\RequirePackage{tikz}
\documentclass[pdflatex,sn-aps,iicol,a4paper]{sn-jnl}

\usepackage{caption}
\usepackage{float}
\usepackage{subcaption}
\usepackage{commath}
\usepackage[T1]{fontenc}
\usepackage[utf8]{inputenc}
\usepackage{xspace}

\def\pythia{\textsc{Pythia}8\xspace}
\def\pyt{\textsc{Pythia}\xspace}
\def\mrm#1{\mathrm{#1}}

\def\ee{\ensuremath{\mrm{e}^+\mrm{e}^-}\xspace}
\def\pp{\ensuremath{\mrm{pp}}\xspace}
\def\pPb{\ensuremath{\mrm{pPb}}\xspace}
\def\PbPb{\ensuremath{\mrm{PbPb}}\xspace}
\def\XeXe{\ensuremath{\mrm{XeXe}}\xspace}
\def\pA{\ensuremath{\mrm{p}A}\xspace}
\def\AA{\ensuremath{AA}\xspace}
\def\NN{\ensuremath{NN}}
\def\eg{\emph{e.g.}\xspace}
\def\ie{\emph{i.e.}\xspace}

\def\ADS{\ensuremath{\delta b_c}\xspace}
\def\SU#1{\ensuremath{SU(#1)}\xspace}
\def\setting#1{\texttt{\footnotesize #1}}

\jyear{2023}%

\begin{document}
\title[~]{A spatially constrained QCD colour reconnection in \pp, \pA, and \AA\ collisions in the \pythia/Angantyr model}

\author[]{\fnm{Leif} \sur{L{\"o}nnblad}}\email{leif.lonnblad@hep.lu.se}

\author[]{\fnm{Harsh} \sur{Shah}}\email{harsh.shah@hep.lu.se}

\affil[]{\orgdiv{Dept.\ of Physics}, \orgname{Lund University}, \orgaddress{\street{S{\"o}lvegatan\ 14A}, \city{Lund}, \postcode{SE-223~62}, \country{Sweden}}}

\abstract{We present an updated version of the QCD-based colour reconnection
  model in \pythia, where we constrain the range in impact parameter
  for which reconnections are allowed. In this way, we can introduce
  more realistic colour reconnections in the Angantyr model for heavy
  ion collisions, where previously only reconnections within separate
  nucleon sub-collisions have been allowed. We investigate how the new
  impact parameter constraint influences final states in \pp
  collisions, and retune parameters of the multi-parton interaction
  parameters in \pyt to compensate so that minimum bias data are
  reproduced. We also study multiplicity distributions in \pA\
  collisions and find that, in order to counteract the loss in
  multiplicity due to the introduction of global colour reconnections,
  we need to modify some parameters in the Angantyr model while keeping the parameters tuned to \pp fixed. With Angantyr we
  can then extrapolate to \AA\ collisions without further parameter
  tuning and retaining a reasonable description of the basic multiplicity
  distributions.}





\maketitle

\section{Introduction}
\label{S:intro}

The field of heavy-ion (HI) collisions is widely studied under
assumption of the creation of a thermalised medium of strongly coupled
partons; the Quark-Gluon-Plasma (QGP).  Observables showing, \eg,
\textit{strangeness enhancement}, long range \textit{collectivity},
\textit{quarkonia suppression}, and \textit{jet quenching} in HI
collisions are conventionally assumed to reflect the formation of such
a QGP~\cite{qgp}.  Two of these observables, namely
\textit{strangeness enhancement}~\cite{strange_17}, and long range
\textit{collectivity}~\cite{ridge}, are, however, observed also in \pp
collisions, where the QGP formation is conventionally not assumed to
be present. This has cast doubts on our understanding of \pp\
collisions as well as on our understanding of these observables
in HI collisions.

\pyt \cite{Pythia,Pythia2} is a well known and widely used event
generator for small collision systems such as \ee and \pp. In
\cite{Angantyr} we developed the Angantyr model to allow the use of
\pyt{'s} excellent description of \pp\ collisions also for HI, by
introducing a sophisticated stacking of multiple \pp-like collisions
to build up complete HI collision events. In this way, we have
constructed a test bench for developing models for collective effects
that can explain the behaviour of the above-mentioned observables
without introducing a QGP.

Angantyr uses an advanced Glauber model \cite{Glauber1, Glauber2},
where the Good--Walker picture \cite{GW} of diffraction provides a
description of fluctuations (often referred to as Glauber--Gribov
colour fluctuations) that influences both the number of
nucleon-nucleon (\NN) sub-collisions and the type of each individual
sub-collision in a heavy-ion event. Each of these sub-collisions is
then simulated using the standard \pyt minimum-bias machinery,
producing non-diffractive, elastic, single and double diffractive
sub-events according to the type determined in the Glauber
simulation. The resulting sub-events are then simply stacked together
into a full HI event.

In Angantyr there is a special treatment of situations where one
nucleon collides non-diffractively with several others. In this
situation only one such sub-collision is considered \textit{primary} and
is modelled by a full non-diffractive event in \pyt. The other,
\textit{secondary}, \pp\ collisions are treated as diffractive excitations
of the additional nucleon and modelled in \pyt using the standard
Pomeron-based single diffractive model. In such a secondary
non-diffractive (SND) sub-collision there is a special treatment of the
Pomeron parton densities, to better mimic the particle production of a
non-diffractive \pp\ event in the direction of the excited nucleon.

The Angantyr model contains a number of new parameters, however most
of these are tuned using \pp observables, except the ones controlling
the SND sub-events, which are tuned to \pA\ observables. It is an
important feature of the Angantyr model that there is then no further
parameters to tune when generating \AA collision events.  The Angantyr
model is nevertheless able to reproduce general features such as
multiplicity distributions in different centrality bins for, \eg, \PbPb
and \XeXe collisions at the LHC.

\begin{figure}
\centering
  \includegraphics[width=\linewidth]{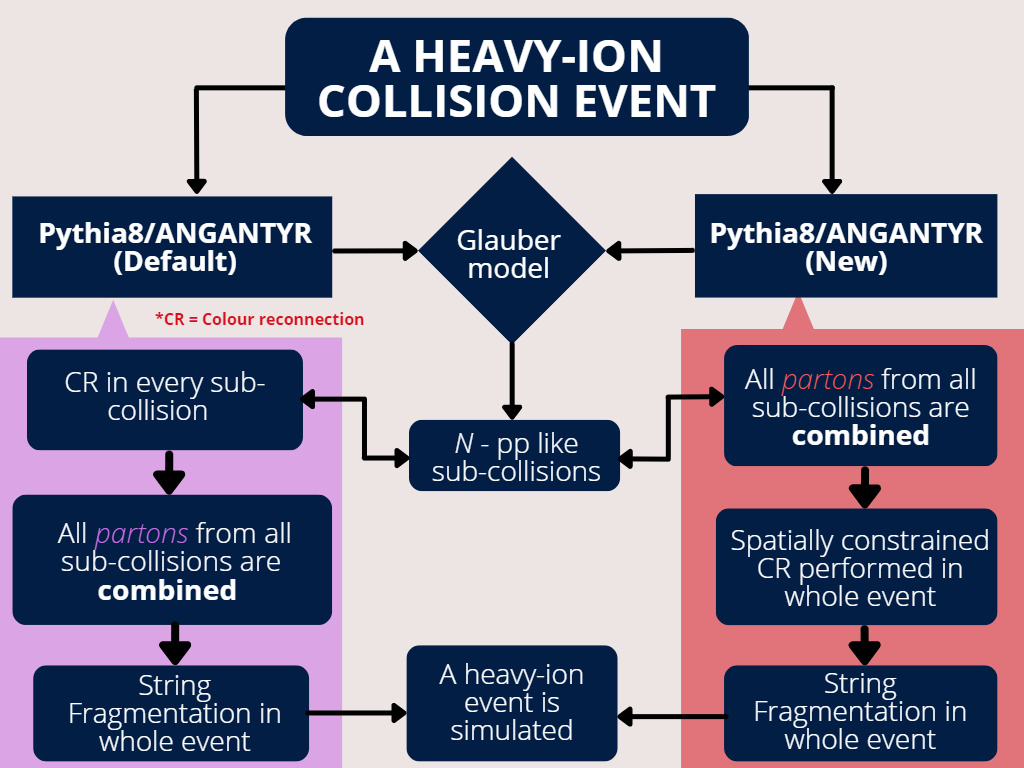}
  \caption{Comparison of the new implementation against the default
    structure of the event simulation in the Angantyr model. So far we
    were treating a heavy-ion collision event as a sophisticated
    superposition of multiple \pp like collisions stacked together at
    the parton level after the colour reconnection. In this work, we
    are treating a heavy-ion event as one event as early as possible,
    which is from the colour reconnection state onward.}
\label{fig:CRO}
\end{figure}

Currently, although the sub-events are generated on parton-level and
stacked together to be hadronized together using \pyt string
fragmentation, the colour dipoles that build up the string from
different sub-collisions do not interact with each other in the
Angantyr model. Therefore in the Angantyr all sub-collisions
hadronize separately in HI events.  Figure \ref{fig:CRO} outlines the
general scheme of the existing HI events simulation in the left part
under \pythia/Angantyr (Default). This is, of course, a simplification
and we don't believe that there is no cross-talk at all between the
sub-events.

This work is aimed to further develop the Angantyr model to have
interactions among partons produced in different sub-collisions in HI
events. Recently two models based on \textit{string interaction} have
been investigated by the Lund group. One is the shoving model
\cite{shoving,shoving2}, where the overlapping fields of nearby
strings give a repulsive force that gives rise to an azimuthal
flow. The other is the rope hadronization model~\cite{rope1, rope2}
where the increased tension in overlapping strings gives rise to
strangeness enhancement. In addition there is now a model for hadron
rescattering in \pythia \cite{hadrescat,hardrescat2}, that also works
for HI collisions.

Here we will instead focus on how the strings are formed from the
coloured partons produced in the scattering and after initial- and
final-state parton showers. This is typically done using the
$N_c\to\infty$ limit where any coloured parton is uniquely coupled to
an anti-coloured one in a \textit{dipole}. Gluons carry both colour
and anti-colour, so we will get a set of dipoles connected together
with gluons that form strings. 

\pyt treats multi-parton interactions (MPIs) \cite{mpi} as independent
partonic interactions, where the initial- and final-state parton
showers are also independent vacuum radiations. Before the produced
strings are allowed to hadronize, however, it was early on clear that
the strings in these different parton interactions needed to
undergo a \textit{colour reconnection} (CR) procedure in order to
describe \pp data.

Colour reconnections is the only step in \pyt where the produced
partons from different sub-scatterings interact with each other before
hadronization. In that spirit, we will here look at the effects of
allowing CR to work also on partons from different \NN\ sub-collisions
in HI collisions as illustrated on the right side of Figure
\ref{fig:CRO} under \pythia/Angantyr (new).

In this work we use the QCD-CR model \cite{QCDCR}, which is different
from the MPI-based CR model \cite{mpi}, which is the default CR model
in \pythia.  We give a short overview of the two CR models in
section \ref{S:cr}.  We introduce a new parameter to determine the allowed spatial
transverse separation between colour dipoles to be colour
reconnected.  This new parameter plays a key role in enabling CR
among partons from different sub-collisions.  We describe its
importance and expected effects on the event final states in more
detail in section \ref{S:Sc2r}.

The rest of this paper is organised as follows. In section
\ref{S:effpara} we discuss the re-tuning of selected parameters needed
for the reproduction of \pp collision data to remain intact when we
introduce the transverse separation cut. There we also describe the
subsequent re-tuning of SND parameters to reproduce multiplicity
distributions in \pPb\ data. We also introduce some modifications to
the fragmentation of so-called junction strings in \pythia (explained in
more detail in appendix \ref{S:jun_frag}), which were necessary to
allow QCD-CR in HI collisions. In section \ref{S:results} we present
some outcomes of the new model in \pp, \pA, and \AA, before we present
our conclusions in section \ref{S:discussion} together with an
outlook.

\section{The Colour Reconnection}
\label{S:cr}

In hadronic collisions there are coloured particles in both the
initial and final state, and it is reasonable to assume that there
will be multiple parton scattering in a single collision. Assuming
that all scatterings are completely independent of each other, and
also contribute equally to the momentum distribution and multiplicity
of final state hadrons, one would expect that the multiplicity would
grow with the number of scatterings, while observables such as the
average transverse momentum, $\langle p_\perp\rangle$, would be almost
constant.

The fact that already the ISR
\cite{Ames-Bologna-CERN-Dortmund-Heidelberg-Warsaw:1983gbi} and UA1
\cite{UA1:1989bou} experiments found that $\langle p_\perp\rangle$
actually increases with the number of charged particles, $N_{ch}$,
tells us that this picture of MPIs is too naive, and indicates some
sort of collective behaviour in the hadronization stage, that
correlates partons from different scatterings. In particular it would
indicate that additional scatterings would contribute to the average
transverse momentum but not so much to the multiplicity.

In the original MPI model \cite{mpi} this was handled with a colour
rearrangement. A single partonic scattering would produce colour
connections between scattered partons and the hadron remnants,
resulting in long strings that produce many (soft) hadrons. A
secondary scattering would naively do the same, but the rearrangement
allows the secondary scattered partons to instead be connected to the
previous scatterings. This reduces the number of additional soft
hadrons produced in additional scatterings. The scatterings will,
however, still contribute to the average transverse momentum, giving a
rise in $\langle p_\perp\rangle$ with multiplicity.

Later the effects of colour rearrangements were also studied in
particle event generators in a series of papers
\cite{Gustafson:1988fs, Sjostrand:1993hi, Gustafson:1994cd,
  Lonnblad:1995yk, Friberg:1996xc} to investigate possible CR in \ee
annihilation at LEP. One of the problems was understanding the
uncertainty in the $W$ mass as measured in $e^+e^- $
$\rightarrow W^+W^-\rightarrow q_1\bar{q_1}'q_2\bar{q_2}'$ events. The
naive expectation is here that the $q\bar{q}$ from each W-decay are
colour connected separately. However, it is possible to have an
alternative configuration where quarks and anti-quarks originating
from different W bosons are colour connected as the final colour
configuration. The difference in colour connections will influence the
jet shapes, and hence also the experimentally reconstructed W
masses. The probability for such a rearrangement of the colour
configuration is given by $1/\mathrm{N^2_c} = 1/\mathrm{9}$.  It is
referred as $\textit{colour reconnection}$ in the event generators.
Today, in event generators such as \pyt, CR is a generic name given to
algorithms which decide a colour configuration to be used to colour
connect the partons.  The probability for CR in lepton collisions is
further reduced by the limited space-time overlap between the produced
parton systems from the W bosons decays in the case of the above
example, as described in \cite{Sjostrand:1993hi}.

Due to the non-perturbative nature and limited understanding of the
colour configuration at the parton level, there is some liberty in
developing CR models.  The common approach in all of them is
minimizing the rapidity span of the produced hadrons from the strings.
For more details, we refer to \cite{mpi, QCDCR} and references therein.

\begin{figure*}
\centering
\begin{tikzpicture}
\draw (-1,3) -- (2,3);
\filldraw [gray] (-1,3) circle (2pt) node[left]{$\bar{q}$};
\filldraw [gray] (2,3) circle (2pt) node[right]{$q$};
\draw (-1,4) -- (2,4);
\filldraw [gray] (-1,4) circle (2pt) node[left]{$q$};
\filldraw [gray] (2,4) circle (2pt) node[right]{$\bar{q}$};

\draw [-stealth](3.2,3.5) -- (4.8,3.5);
\filldraw [gray] (6,3) circle (2pt) node[left]{$\bar{q}$};
\filldraw [gray] (9,3) circle (2pt) node[right]{$q$};
\filldraw [gray] (6,4) circle (2pt) node[left]{$q$};
\filldraw [gray] (9,4) circle (2pt) node[right]{$\bar{q}$};
\draw (6,3) -- (6,4);
\draw (9,3) -- (9,4);

\node[] at (4,2.5) {(a)};

\draw (-2,2) -- (10, 2);

\draw (-1,0) -- (2,0);
\filldraw [gray] (-1,0) circle (2pt) node[left]{$q$};
\filldraw [gray] (2,0) circle (2pt) node[right]{$\bar{q}$};
\draw (-1,1) -- (2,1);
\filldraw [gray] (-1,1) circle (2pt) node[left]{$q$};
\filldraw [gray] (2,1) circle (2pt) node[right]{$\bar{q}$};

\draw [-stealth](3.2,0.5) -- (4.8,0.5);

\draw (6,0) -- (6.5,0.5);
\filldraw [gray] (6,0) circle (2pt) node[left]{$q$};
\draw (6,1) -- (6.5,0.5);
\filldraw [gray] (6,1) circle (2pt) node[left]{$q$};

\draw (6.5,0.5) -- (8.5, 0.5);

\draw (9,0) -- (8.5,0.5);
\filldraw [gray] (9,0) circle (2pt) node[right]{$\bar{q}$};
\draw (9,1) -- (8.5,0.5);
\filldraw [gray] (9,1) circle (2pt) node[right]{$\bar{q}$};
\node[] at (4,-0.5) {(b)};

\draw (-2,-1) -- (10, -1);
\draw (-1,-2) -- (2,-2);
\filldraw [gray] (-1,-2) circle (2pt) node[left]{$q$};
\filldraw [gray] (2,-2) circle (2pt) node[right]{$\bar{q}$};
\draw (-1,-3) -- (2,-3);
\filldraw [gray] (-1,-3) circle (2pt) node[left]{$q$};
\filldraw [gray] (2,-3) circle (2pt) node[right]{$\bar{q}$};
\draw (-1,-4) -- (2,-4);
\filldraw [gray] (-1,-4) circle (2pt) node[left]{$q$};
\filldraw [gray] (2,-4) circle (2pt) node[right]{$\bar{q}$};

\draw [-stealth](3.2,-3) -- (4.8,-3);

\draw (6,-2) -- (7,-3);
\filldraw [gray] (6,-2) circle (2pt) node[left]{$q$};
\draw (6,-3) -- (7,-3);
\filldraw [gray] (6,-3) circle (2pt) node[left]{$q$};
\draw (6,-4) -- (7,-3);
\filldraw [gray] (6,-4) circle (2pt) node[left]{$q$};

\draw (9,-2) -- (8,-3);
\filldraw [gray] (9,-2) circle (2pt) node[right]{$\bar{q}$};
\draw (9,-3) -- (8,-3);
\filldraw [gray] (9,-3) circle (2pt) node[right]{$\bar{q}$};
\draw (9,-4) -- (8,-3);
\filldraw [gray] (9,-4) circle (2pt) node[right]{$\bar{q}$};
\node[] at (4,-4) {(c)};

\end{tikzpicture}
\caption{Two dipoles and three dipoles CR possibilities.  For two
  dipoles, they can either have (a) a simple reconnection (a.k.a.\ a
  \textit{swing}) or (b) a formation of a connected junction and
  anti-junction system.  Three dipoles can form (c) disconnected
  junction and anti-junction systems.}
\label{fig:dipCR}
\end{figure*}
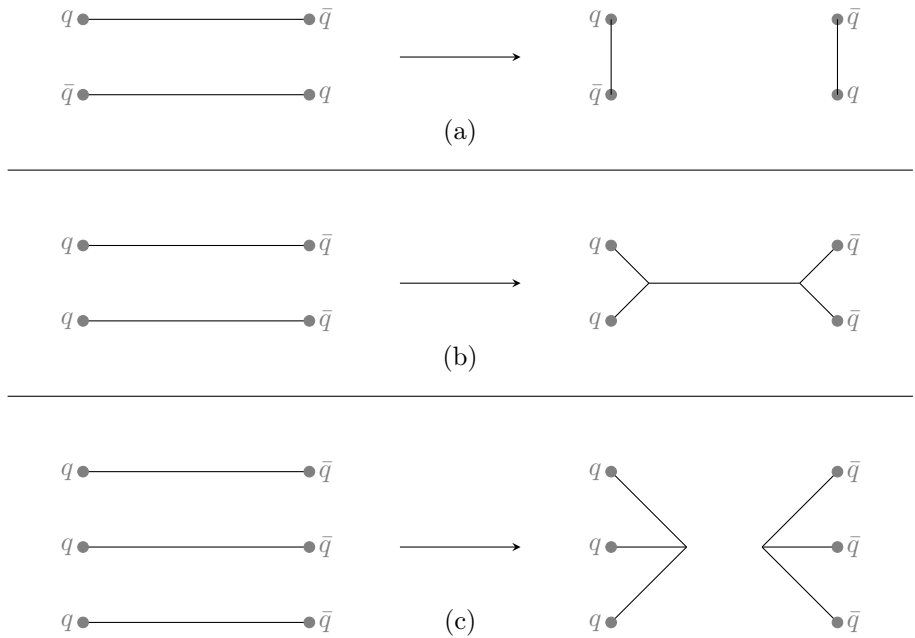

The possibility of a CR model based on \SU{3} colour algebra with a
finite number of colours was first proposed in \cite{Gustafson:1988fs}
for the W pair production and their purely hadronic decays in \ee
collisions.  The QCD-CR model \cite{QCDCR} is an extension of the
default CR in \pyt introducing \SU{3} colour algebra in \pp
collisions.  In addition to the standard reconnection of colour lines
(sometimes referred to as a \textit{swing}, see Figure
\ref{fig:dipCR}a), this model also includes the possibility for
$\textit{junction}$ formation, where three string pieces are connected
to a single point. Junctions can be formed by two dipoles reconnecting
into a junction--anti-junction pair connected by a new colour line as
in Figure \ref{fig:dipCR}b, or by three dipoles reconnecting to
separate junction---anti-junction systems as in Figure
\ref{fig:dipCR}c. As the junctions carry baryon number, the QCD-CR
model introduces a new baryon production mechanism, in addition to the
normal formation of baryons in the string fragmentation through
diquark production.

Before we discuss junctions and other possibilities for colour
reconnection under QCD-CR, let us see what makes it beyond the leading
colour (LC).  Referring to the arguments in\cite{QCDCR}, consider a
scenario when two gluons are extracted from a proton, here QCD gives
several possibilities for the colour multiplets formed by these two
gluons.  Each of the gluons can have $8$ possible colours, and from
the colour algebra, these two gluons are in one of the colour
multiplets,
\[
8 \otimes 8 = 27 \oplus 10 \oplus \bar{10} \oplus 8 \oplus 8 \oplus 1.
\]
The $27$ (or a "viginti-septet") represents LC, where the two gluons
extend four independent strings from the proton. For random gluon
colours, this has the probability
\[
P_{LC} = \frac{27}{64} \approx 0.4,
\]
which is less than 50\%.  Hence it is evident that sub-leading colour
topology has non-negligible effects.  We note that the least probable
colour configuration is $1$ (singlet), where both gluons have exactly
opposite colours, and it can occur with a probability of $1/64$.
Similarly, other possible cases of two quarks, a quark and a gluon, or
a quark and an anti-quark are shown in \cite{QCDCR}.  The conclusion
is that sub-leading multiplets are ignored in a LC model, and
for hadron collisions, sub-leading multiplets have a significant
contribution.  The colour algebra becomes more and more complex for
cases where multiple partons are extracted from the beam particle.

The QCD-CR starts with LC ($N_c \rightarrow \infty$) connections after
the parton showers and assigns \SU{3} weighted colours to the
partons.  The above-mentioned \SU{3} colour rules are applied only on
the uncorrelated partons.  The assumption of uncorrelated partons, and
assigning them \SU{3} weighted colour compositions allows us to have
an approximation of the colour configuration the partons may have.
Once all the partons are assigned colours (one colour for a quark, and
two colour indices for a gluon), the colour reconnection is performed
based on the fundamental idea of minimising the so-called
$\lambda$-measure \cite{lambda:1989}.  The $\lambda$-measure gives an
estimate of the number of hadrons produced in the string breakup.

Three different initial colour topologies and respectively allowed
reconnections are shown in figure \ref{fig:dipCR}, for the colour
dipoles between a quark and an anti-quark.  There is one more
configuration for dipoles with a long gluon chain, which are colour reconnected as
$\textit{zipper-style}$ junctions, but we are not going into details
about such complex configurations here.

For two colour dipoles, there are two possible reconnection
topologies:

\begin{itemize}
\item[(a)] Ordinary style (Colour dipole swing): In this case, both
  dipoles exchange their colour connected partons.  The QCD colour
  algebra will control the reconnection probability in addition to the
  $\lambda$ measure for the new configuration.  In this particular
  case, the reconnection probability from the QCD constraints is
  $1/9$, since both dipoles have to have the same colour
  configuration.
\item[(b)] Junction style: Instead of the reconnection of the dipole
  endpoints, a new string piece is created connecting the two quarks
  to one end of the string piece, and two anti-quarks to the other
  end.  This configuration creates a junction and an anti-junction
  connected by a string piece.  The QCD probability is here $1/3$,
  which is higher than in the previous case.  However, the potential
  reduction in the $\lambda$-measure in this type of configuration is
  smaller, due to the creation of a new string piece.  Hence, the
  algorithm suppresses such a configuration.
\end{itemize}
The QCD-CR model has also a possibility for reconnection for three colour dipoles:
\begin{itemize}
\item[(c)] Junction style (Three dipoles): Two independent string
  systems with a junction and an anti-junction are formed after the
  reconnection.  In this case, the QCD probability is $1/27$.
\end{itemize}

Similarly, colour reconnections are performed for dipoles containing
$q - g, g - g$, and $\bar{q} - g$.  For all types of colour dipoles,
the reconnection probability is controlled by the \SU{3} colour algebra
and the $\lambda$-measure.

In the model implementation, some simplifications are made.  All
dipoles are assigned colour indices from $1$ to $9$.  For the
ordinary swing reconnection to be allowed, two dipoles have to have
the same index, which will provide a $1/9$ probability.  For the case
of junction style reconnection between two dipoles, the constraint is
that the dipole indices have to be different but their value modulo
three has to be the same, giving the probability $2/9$. Similarly, for
the three dipoles case the junction style CR will require all three
dipoles to have different indices, but the same value for the modulo
three of the dipole indices, giving the probability $2/81$.

Clearly the junction formation reconnections are here suppressed
compared to the pure colour algebra ($1/3\to2/9$ and $1/27\to2/81$
respectively), and to compensate for this a special parameter $C_j$ is
used to decrease the $\lambda$-measure for junction systems in the
model to favour junction formation over the swing reconnection.

The above constraints will only decide if a certain colour
configuration will be allowed or not.  To determine if the allowed
configuration is preferred or not, the model calculates the
$\lambda$-measure.  The model only allows the reconnection between the
two or three dipoles if the $\lambda$-measure of the new configuration
is lower than the original one.

\subsection{Spatially constrained model}
\label{S:Sc2r}

In the QCD-CR model, all colour dipoles are allowed to undergo CR in
\pp collisions in \pyt.  For any two (or three) dipoles, whether they
will be colour reconnected or not, is primarily decided based on
whether the colour indices match and whether or not the colour
reconnection will reduce the overall $\lambda$-measure.

Our aim is to treat super-positioned \pp like events as a single HI
event as early as possible in the Angantyr model.  One way is to
perform CR on all the colour dipoles from all the sub-collisions
before the hadronization stage in the Angantyr model.  But the spatial
span of a HI collision can be as large as the diameters of the two
colliding nuclei.  The strong force is a short range force, and its
range is approximately the size of a proton.  Therefore, it is
essential to introduce an additional spatial constraint between the
colour dipoles to be colour reconnected.

The primary assumption in this work is that the majority of the colour
dipoles are more or less parallel to the beam axis, and they are
separated in the transverse plane.  Hence we use the transverse
positions of the partons to determine the separation between any two
colour dipoles.  For a more realistic constraint, one must take into
account the full 3+1-dimensional space-time coordinates. Here we
instead make a simplified assumption that the position of the dipole
in the impact parameter can be represented by the midpoint between the
two partons. If the distance between two such dipole's midpoints is
larger than some parameter, $\delta b_c$, they should not be allowed
to reconnect.

When we apply the spatial constraint in the QCD-CR model in \pp
collisions, the direct consequence of the constraint will be on the
multiplicity distribution: fewer dipoles are now allowed to undergo CR
compared to the default setup, the total multiplicity distribution
will increase. But also other observables will be affected, such as
$\langle p_\perp \rangle (N_{ch})$ and the $p_\perp$ distribution. Before
applying the spatial constraint to HI collisions, we will therefore
need to re-tune some of the parameters of the MPI and QCD-CR models to
retain a good description of \pp\ data.

In HI collisions, the spatial constraint will allow CR among the
nearby colour dipoles independent of their original sub-collisions.
This will increase CR in a HI event, especially in $\textit{central}$
collisions, and thus reduce the multiplicity. To counteract this we will
need to retune some of the parameters in the Angantyr model.

In section \ref{S:effpara}, we will discuss the selection of the
parameters we will retune and examine how they affect relevant
observables.

\subsection{Improved junction handling}
\label{sec:impr-junct-handl}

Compared to the default reconnection model in \pythia, the QCD-CR
results in an increase in simulation time. This could be expected due
to the increase in complexity of the algorithm. We noticed, however,
that the primary cause of the increase was the high failure rate in
the hadronization of the junction systems. Many junction systems do not
hadronize properly at the first attempt, and the algorithm has to
repeat the process multiple times to succeed.

Moreover, when running the QCD-CR for HI collisions we noticed that a
substantial number of events were thrown away because \pythia was not
able to hadronize certain complicated junction systems. Normally such
errors are unimportant, but since these errors were more frequent in
high multiplicity events, they caused an artificial skewing of the
overall multiplicity distribution. This is noticeable already in
\pp and became quite significant in \AA. For this reason, we decided
to improve the handling of the junction hadronization in \pythia. The
changes we made are mainly technical improvements and will be
included in a future \pythia release. For completeness, we include the
details of these changes in appendix \ref{S:jun_frag}.

\begin{figure*}
\centering
\begin{subfigure}{.5\textwidth}
  \includegraphics[width=\linewidth]{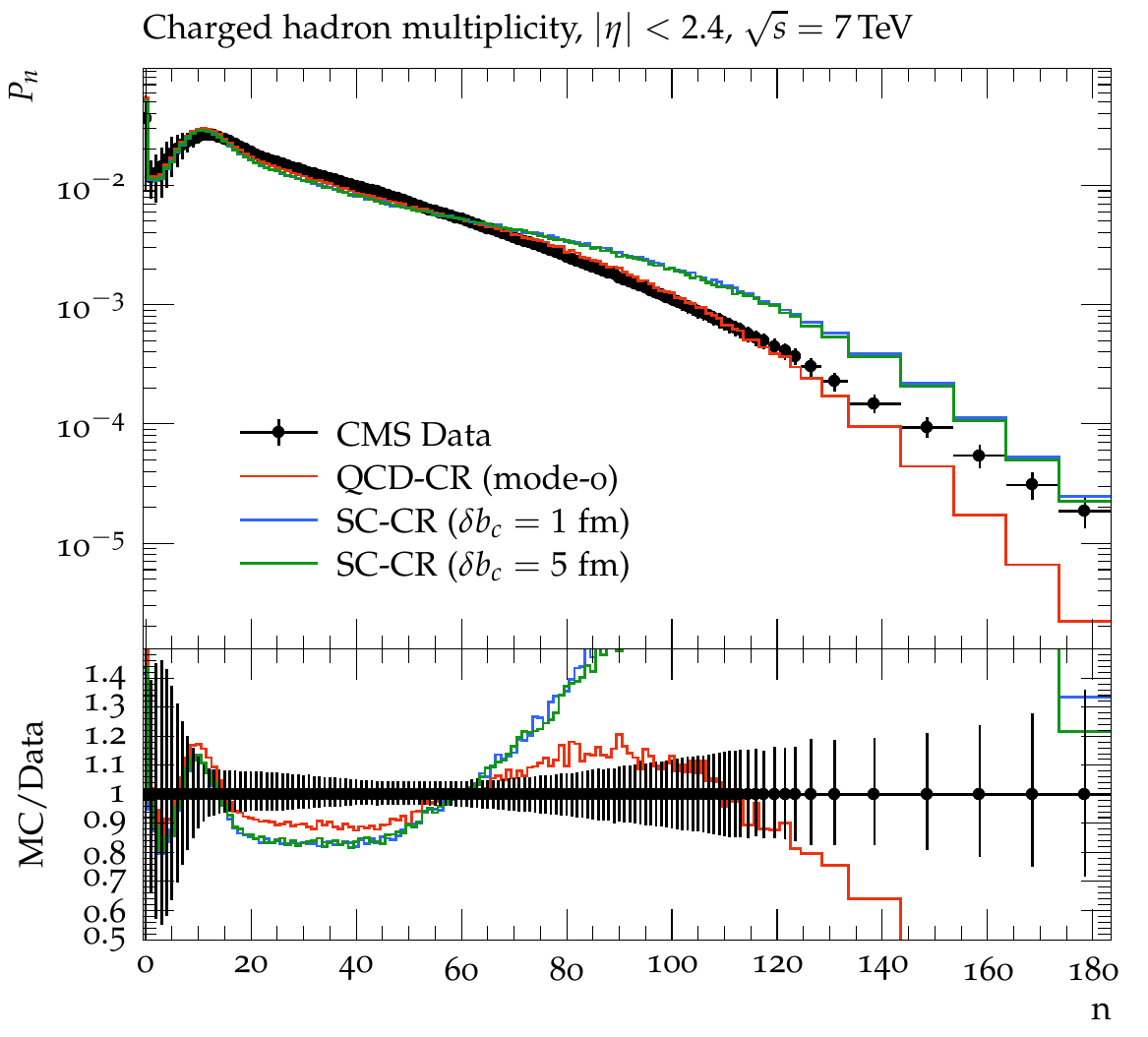}
\end{subfigure}%
\begin{subfigure}{.5\textwidth}
  \includegraphics[width=\linewidth]{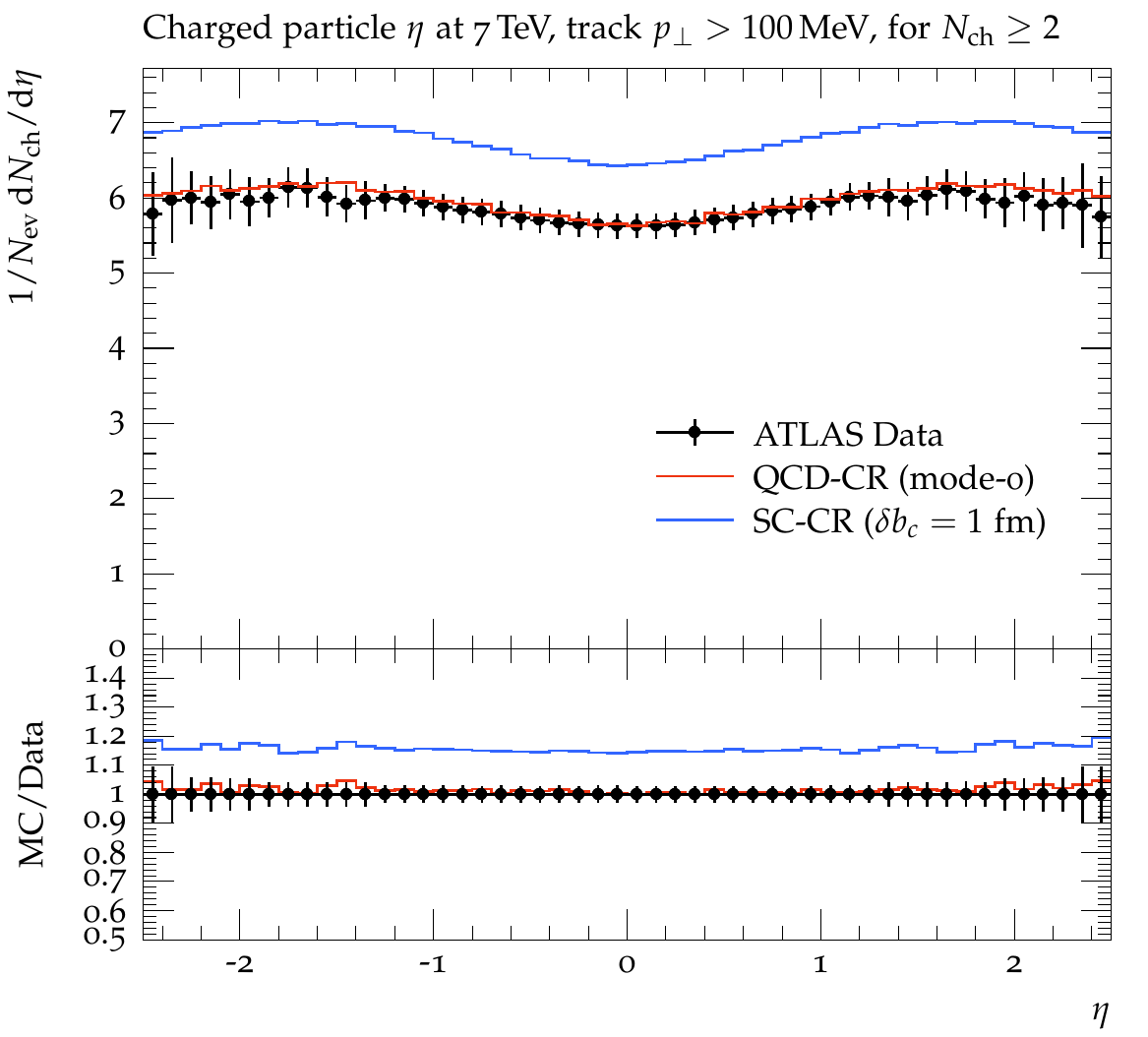}
\end{subfigure}
\caption{The effect of improved junction hadronization in
  \pythia before retuning. Events are generated for $\sqrt{s}=7$~TeV \pp
  non-single-diffractive collisions and compared with CMS
  \cite{CMS:2010qvf} and minimum bias collisions with ATLAS
  \cite{ATLAS:2010jvh} data. $\textit{Left}$: multiplicity in $\mid \eta \mid$ <
  2.4 with CMS data. $\textit{Right}$: multiplicity distribution in $\eta$ for
  $\mid \eta \mid$ < 2.4 with ATLAS data. The red line shows the
  results for default QCD-CR (Mode-0) in \pythia, and the blue and
  green lines represent the results for SC-CR with improved junction
  hadronization and added spatial constraint set to large values: blue
  lines $\delta b_c=1$~tm and green line $\delta b_c=5$~fm. }
\label{fig:pp_prim_mult}
\end{figure*}

The direct consequence of our modifications in the junction
hadronization can be seen in \pp\ collisions as an enhancement of high
multiplicity events.  In figure \ref{fig:pp_prim_mult} we show example
multiplicity distributions in CMS and ATLAS minimum bias events,
compared to the result from the QCD-CR(mode-0) predictions and the
results of our modifications (labelled SC-CR for \textit{Spatially
  Constrained} CR), but with an allowed dipole separation so large
that only the effects of the modified junction hadronization are
included. In the leftmost figure, we show two values of the allowed
dipoles separation, $\delta b_c=1$ and $5$~fm, to show that already
1~fm is large enough to remove the effect of the spatial constraints.

Clearly, the QCD-CR needs to be returned when introducing the improved
junction hadronization, but in the following, we will also want to tune
the value of the allowed dipole separation in light of using the SC-CR
model also for heavy ion collisions.

\section{Selection of parameters and retuning strategy}
\label{S:effpara}

There are many parameters within and outside the QCD-CR model, which
can be re-tuned.  Table \ref{tab:table1} shows the set of the
parameters re-tuned when the QCD-CR model is introduced in \pyt.
\begin{table*}[!htb]
\centering
\begin{tabular}{l*{2}{c}}
  Parameters & Monash & QCD-CR (Mode-0) \\
  \hline
  \setting{StringPT:sigma} & 0.335 & 0.335 \\
  \setting{StringZ:aLund} & 0.68 & 0.36 \\
  \setting{StringZ:bLund} & 0.98 & 0.56 \\
  \setting{StringFlav:probQQtoQ} & 0.081 & 0.078 \\
  \setting{StringFlav:ProbStoUD} & 0.217 & 0.2 \\
  \setting{StringFlav:probQQ1toQQ0join} & 0.5, 0.7, 0.9, 1.0 & 0.0275,0.0275,0.0275,0.0275 \\
  \hline
  \setting{BeamRemnants:remnantMode} & 0 & 1 \\
  \setting{BeamRemnants:saturation} & -&5 \\
  \hline
  \setting{MultiPartonInteractions:pT0Ref} ($p^{\text{ref}}_{\perp0}$)& 2.28 & 2.12 \\
  \hline
  \setting{ColourReconnection:mode} & 0 & 1 \\
  \setting{ColourReconnection:allowDoubleJunRem} & - & off \\
  \setting{ColourReconnection:m0} ($m_0$) & -&2.9 \\
  \setting{ColourReconnection:allowJunctions} & -&on \\
  \setting{ColourReconnection:junctionCorrection} ($C_j$) & -&1.43 \\
  \setting{ColourReconnection:timeDilationMode} &-& 0 \\
\end{tabular}
\caption{The list of parameters and their values in the Monash tune, and in the QCD-CR (Mode-0) tune in \pythia.}
\label{tab:table1}
\end{table*}

The parameters in the first part of Table \ref{tab:table1} are tuned
outside of the QCD-CR model.  Those parameters directly affect the
flavour production under string breaking during the fragmentation
stage in the hadronization framework. The primary reason for
re-tuning those parameters was to adjust the flavour production under
the new colour reconnection treatment. We decided not to modify the
parameters associated with the string breaking and quark-antiquark
pair creation with the parameters in this analysis, and instead mainly
focus on the parameters governing the overall multiplicity.

We selected the following parameters to retune against \pp\ data:
\begin{enumerate}
\item $p^{\text{ref}}_{\perp0}$, low-$p_\perp$ suppression for MPIs,
\item $m_0$, a scale parameter used in the $\lambda$-measure and a
  mass cut-off for pseudo-particles (see section \ref{mcj}),
\item $C_j$, a parameter reducing the $\lambda$-measure for junction
  systems,
\item  \ADS, the allowed dipole separation.
\end{enumerate}
The last one of these is the new parameter we introduce in
\pyt\footnote{In \pyt this parameter is now called
  \setting{ColourReconnection:dipoleMaxDist}.} which gives maximum
transverse distance between centres of a dipole pair to be considered
for reconnection (in femtometres). \ADS also restricts the junction
formation with three dipoles. If any of the three dipole pairs is
separated with a transverse distance larger than \ADS, then those
three dipoles are not allowed to form junctions.

The primary strategy for our retuning is to first modify these four
parameters, and after an acceptable tune for \pp\ observables has been
obtained we turn to \pA\ and try to adjust parameters in the Angantyr
model to also get an acceptable description there. The main feature of
the Angantyr model affecting the overall behaviour of the multiplicity
is the treatment of secondary non-diffractive (SND) sub-collisions.

The SND interactions are introduced in the Angantyr model to treat
situations, where a nucleon is tagged as a participant in multiple
non-diffractive type collisions with other nucleons. While primary
collisions are generated as normal non-diffractive \pp\ events, these
secondary ones are generated as a diffractive excitation of the
additional nucleon. This is the main feature that allows Angantyr to
reproduce general features of the final state multiplicity in \pA and
\AA events.  In \cite{Angantyr} we modified the Pomeron parton
distribution functions (PDFs) for SND events. In this work, we have
decided to rather modify the Pomeron flux in SND events, and modify
the so-called $\epsilon_{pom}$ parameter. Modification of this
parameter is aimed to modify only the SND interactions. Since SND
interactions do not occur in \pp collisions, we can not tune the
$\epsilon_{pom}$ parameter there.

After obtaining a reasonable fit to \pA\ data, we can use the obtained
tune to generate \AA events, and at this stage, there are no further
parameters to tune.

In the following we will study the effects of the selected parameters
individually. We will restrict our study to typical minimum bias
observables such as charged multiplicity distributions and
$\langle p_\perp \rangle (N_{ch})$.

\subsection{ Low-$p_\perp$ suppression for MPIs}
For minimum bias events, \pyt regularise the QCD $2\rightarrow 2$
process by introducing a collision energy dependent parameter
$p_{\perp0}$ to regularise the divergences in the partonic cross
section
\begin{equation}
  \frac{d\sigma_{2\rightarrow 2}}{dp^{2}_{\perp}} \rightarrow\frac{d\sigma_{2\rightarrow 2}}{dp^{2}_{\perp}} 
  \times\frac{p^4_\perp} {(p^2_\perp + p^2_{\perp0})^2}
  \frac{\alpha^{2}_{s} (p^2_\perp + p^2_{\perp0})}{\alpha^{2}_{s} (p^2_\perp)}
\end{equation}

Here $\alpha_{s}$ is the strong coupling constant, and $p_\perp$ is
the transverse momentum of the scattered partons. The energy
dependence of the parameter $p_{\perp0}$ is further given by another
parameter $p^{\text{ref}}_{\perp0}$. The power law dependent relation
is given as
\begin{equation}
  p_{\perp0} = p_{\perp0}(E_{\text{cm}}) = p^{\text{ref}}_{\perp0} \times
  \left(\frac{E_{\text{cm}}}{E_{\text{cm}}^{\text{ref}}}\right)^{p_{E_{\text{cm}}}},
\end{equation}
where $p_{E_{\text{cm}}}$ is a scaling parameter, controlling the
growth of $p_{\perp0}$ with the centre of mass energy of the
collision, $E_{\text{cm}}$, with respect to a reference energy,
$E_{\text{cm}}^{\text{ref}}$, which by default is set to 7~TeV in
\pyt.

\begin{figure*}
\centering
\begin{subfigure}{.5\textwidth}
  \includegraphics[width=\linewidth]{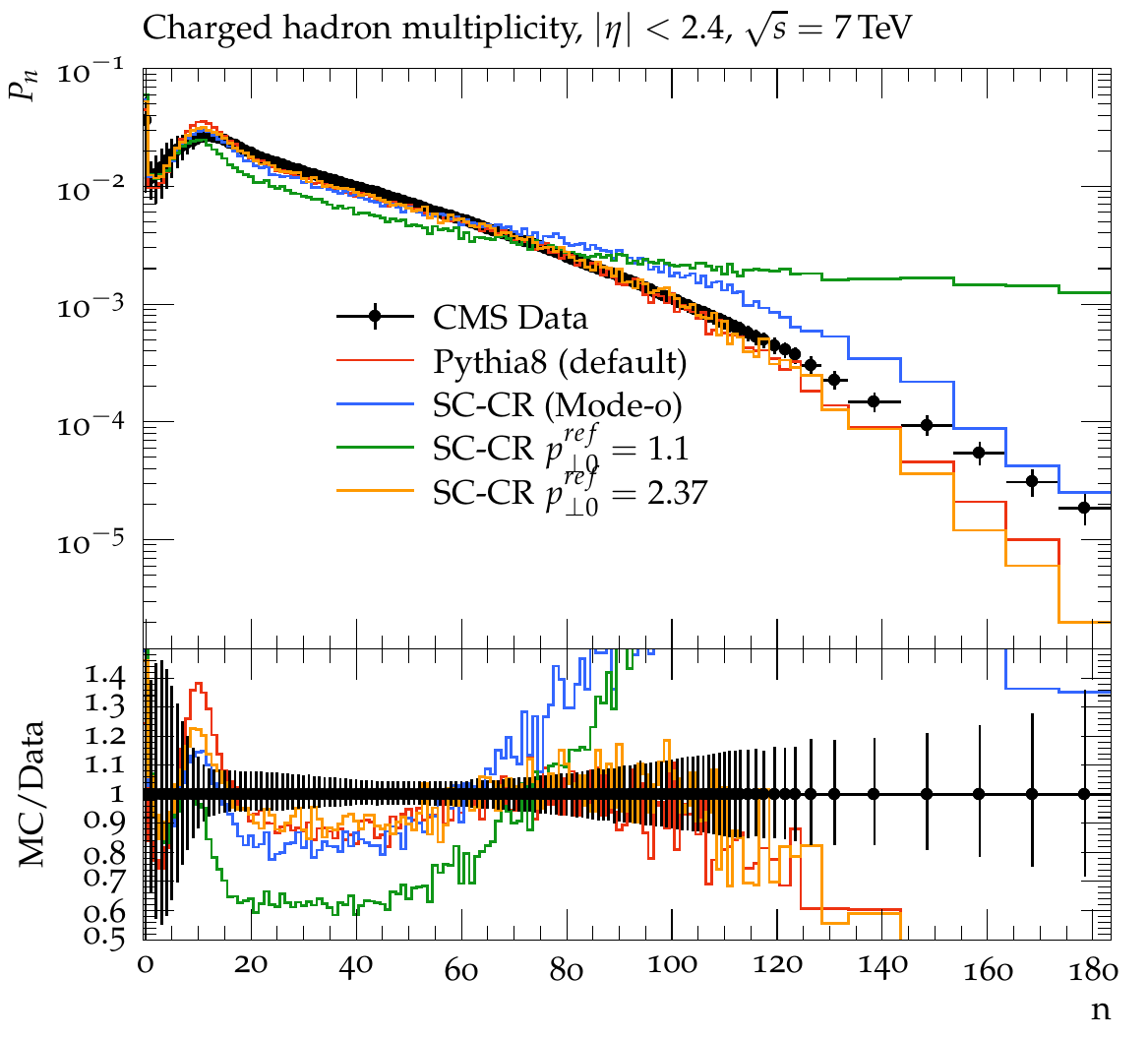}
\end{subfigure}%
\begin{subfigure}{.5\textwidth}
  \includegraphics[width=\linewidth]{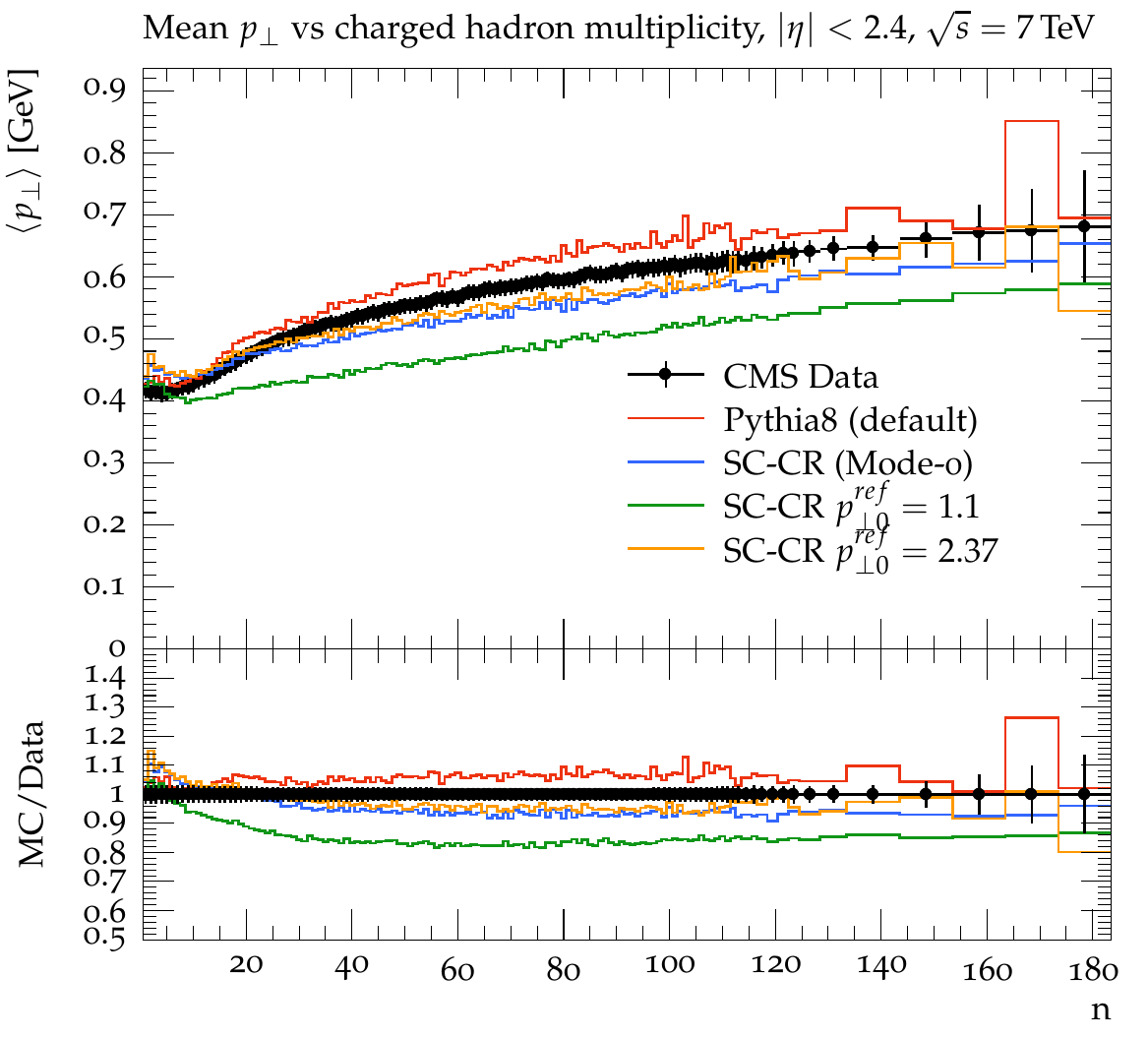}
\end{subfigure}
\caption{The effect of varying the $p^{\text{ref}}_{\perp0}$
  parameter in \pythia. Events are generated for $\sqrt{s}=7$~TeV \pp
  non-single-diffractive collisions and compared with CMS data
  \cite{CMS:2010qvf}. $\textit{Left}$: multiplicity in $\mid \eta \mid$ <
  2.4. $\textit{Right}$: $\langle p_\perp \rangle$ vs $N_{ch}$ for
  $\mid \eta \mid$ < 2.4. The red line shows the results for default
  \pythia, the blue line the results for SC-CR (Mode-0), while the
  green and orange lines show the effects of varying
  $p^{\text{ref}}_{\perp0}$ in the latter.}
\label{fig:pp_ptref2_mult}
\end{figure*}

The effects of varying $p^{\text{ref}}_{\perp0}$ on the event
multiplicity and $\langle p_\perp \rangle (N_{ch})$ distribution is
shown in Figure \ref{fig:pp_ptref2_mult}. Here (and also in Figures
\ref{fig:pp_m02_mult} and \ref{fig:pp_cj2_mult} below) we compare
result for default \pythia\ with the results for the SC-CR (with all
parameters set as in Mode-0 in Table \ref{tab:table1} and with
$\ADS=1$~fm). The main effect of reducing
$p^{\text{ref}}_{\perp0}$ is an increase of multiple scatterings which
increases the multiplicity and reduces the average transverse momenta,
which is indeed what is shown in the figure.

\subsection{$m_0$ and $C_j$ parameters}
\label{mcj}

The parameter $m_0$ is the mass scale in the $\lambda$-measure
\cite{lambda:1989} used in the QCD-CR model.  $C_j$ is a parameter
which modifies this mass scale in string pieces connected to a
junction, according to $m_{0j}= C_jm_0$.

In the QCD-CR model, there is a special treatment of small-mass string
pieces. Any dipole with invariant mass less than the $m_0$ scale will
not be allowed to reconnect but is instead collapsed into a
pseudo-particle.

\begin{figure*}
\centering
\begin{tikzpicture}
\draw (-2,0) -- (-1,1);
\filldraw [gray] (-2,0) circle (2pt) node[left]{q};
\filldraw [gray] (-1,1) circle (2pt) node[right]{g};
\draw (-1,1) -- (-0.3,0.5);
\filldraw [gray] (-0.3,0.5) circle (2pt) node[below]{g};
\draw (-0.3,0.5) -- (1,0.3);
\filldraw [gray] (1,0.3) circle (2pt) node[right]{$\bar{q}$};

\draw [-stealth](2.5,0.5) -- (3.5,0.5);

\draw (5,0) -- (6.2,0.75);
\filldraw [gray] (5,0) circle (2pt) node[left]{q};
\filldraw [gray] (6.2,0.756) circle (2pt) node[left]{(gg)};
\draw (6.2,0.75) -- (8,0.3);
\filldraw [gray] (8,0.3) circle (2pt) node[right]{$\bar{q}$};

\draw (-2,-2) -- (-0.5,-2);
\filldraw [gray] (-2,-2) circle (2pt) node[left]{q};
\filldraw [gray] (-0.5,-2) circle (2pt) node[above]{g};
\draw (-0.5,-2) -- (0,-2.5);
\draw (0,-2.5) -- (1,-2.5);
\filldraw [gray] (1,-2.5) circle (2pt) node[right]{q};
\draw (0,-2.5) -- (-0.5,-3);
\filldraw [gray] (-0.5,-3) circle (2pt) node[below]{g};
\draw (-0.5,-3) -- (-2,-3);
\filldraw [gray] (-2,-3) circle (2pt) node[left]{q};

\draw [-stealth](2.5,-2.5) -- (3.5,-2.5);

\draw (8,-2.5) -- (7,-2.5);
\filldraw [gray] (8,-2.5) circle (2pt) node[right]{q};
\draw (7,-2.5) -- (5,-2);
\filldraw [gray] (5,-2) circle (2pt) node[left]{q};
\draw (7,-2.5) -- (5,-3);
\filldraw [gray] (5,-3) circle (2pt) node[left]{q};
\filldraw [gray] (7,-2.5) circle (2pt) node[above]{(gg)};

\end{tikzpicture}
\caption{Top) Pseudo-particle is formed from a dipole that has a
  smaller invariant mass than $m_0$ in a string, and Bottom)
  Pseudo-particles are formed if the dipole is connected to a junction.}
\label{fig:pseudo}
\end{figure*}
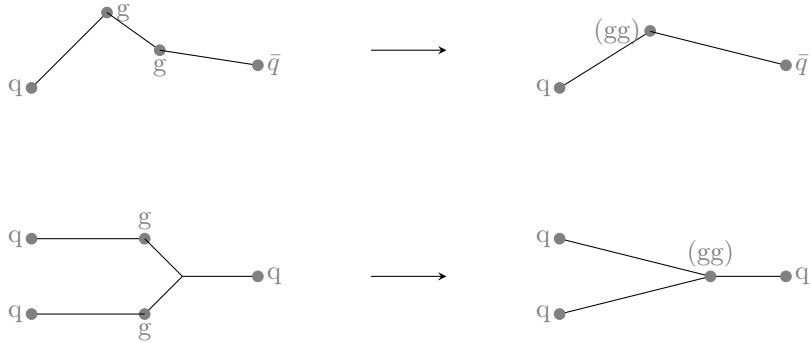

Figure \ref{fig:pseudo} shows how one short dipole or two short
dipoles are replaced by a $\textit{pseudo-particle}$ in a normal
dipole chain (top panel), and in a junction system (bottom panel).
The invariant mass of every dipole is compared with $m_0$, and if it's
smaller than $m_0$ then the dipole is replaced with a
$\textit{pseudo-particle}$, which has the four-momentum of the dipole.
Dipoles on both ends of the short dipole are connected to the
$\textit{pseudo-particle}$.  This way the value of $m_0$ will directly
control the number of dipoles that undergoes CR in the QCD-CR model.
These small dipoles are only removed during the CR, but they
do contribute to the hadron production.  Therefore the parameter $m_0$
significantly affects the hadron multiplicity in the event final
state.  The removal of $\textit{small}$ dipoles is due to technical
reasons, and to reduce the complexity of the CR process \cite{QCDCR}.
It is suggested that in the QCD-CR \cite{QCDCR} model the parameter
$m_0$ having its value around $\Lambda_{QCD}$ has negligible effect,
but raising its value beyond 1~GeV significantly reduces the amount of
the colour reconnection.

The $\lambda$-measure used in \cite{lambda:1989} is an infrared safe
measure of partonic final states, approximately proportional to the
the resulting hadronic multiplicity. In the QCD-CR, an approximation
is used, which is good for large energies. For string pieces between
gluons and/or massless quarks, it is given by
\begin{equation}
    \lambda = ln \left(1 + \frac{\sqrt{2}E_1}{m_0}\right) + ln \left(1 + \frac{\sqrt{2}E_2}{m_0}\right),
\end{equation}
where energies are calculated in the dipole's rest frame.
In the parentheses, $\mathbf{1}$ is added to avoid negative contributions.

In the current QCD-CR implementation in \pythia, the value for
$\textit{pseudo-particle}$ mass cut-off is also used in the
calculation of the $\lambda$-measure. But in principle, one could treat
them as independent parameters.

Eq. (3) is also used for a string piece connected to a junction, where
the energy is measured in the junction rest frame. For a string piece
connecting two junctions the $\lambda$-measure is given by
\cite{lambda:junction}. The distance between the two junctions is
also added to the calculation of the $\lambda$-measure of the new
system and is given by
\begin{equation}
  \lambda = \log \left(\beta_{j1} \beta_{j2} + \sqrt{(\beta_{j1} \beta_{j2})^2 - 1}\right),
\end{equation}
where $\beta_{j1}$ and $\beta_{j2}$ are 4-velocities of the two junctions.

The parameter $m_{0j}$ is used in the $\lambda$-measure calculation for the
junction systems, while the $m_0$ is used for the $\lambda$-measure of
the dipoles. Increasing $m_{0j}$ results in a lower value for the
$\lambda$-measure, which favours junction production. In this way the
suppression of junctions in the model compared to proper \SU{3}
algebra can be compensated by using a value above unity for the
parameter $C_j$.

\begin{figure*}
\centering
\begin{subfigure}{.5\textwidth}
  \includegraphics[width=\linewidth]{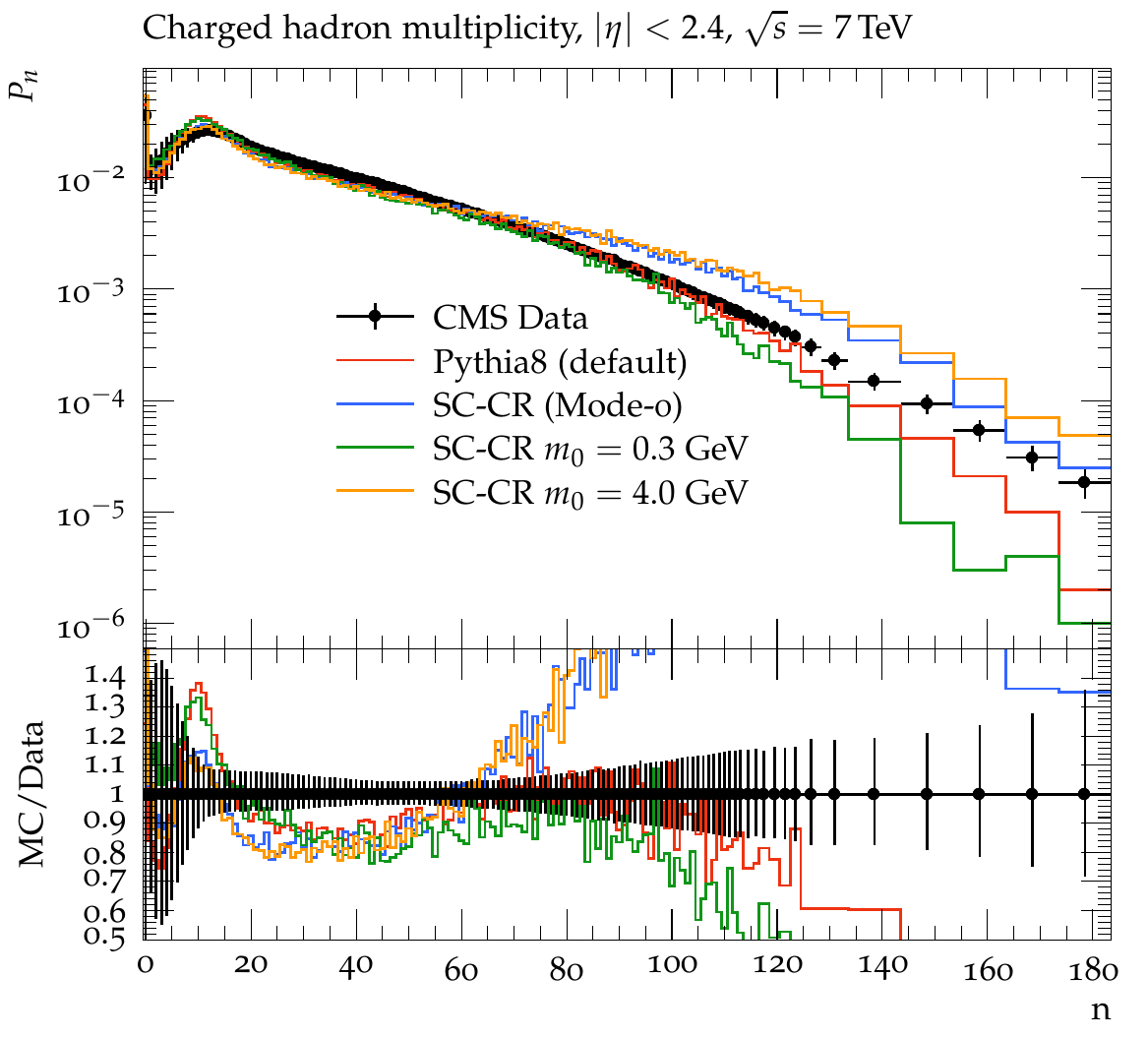}
\end{subfigure}%
\begin{subfigure}{.5\textwidth}
  \includegraphics[width=\linewidth]{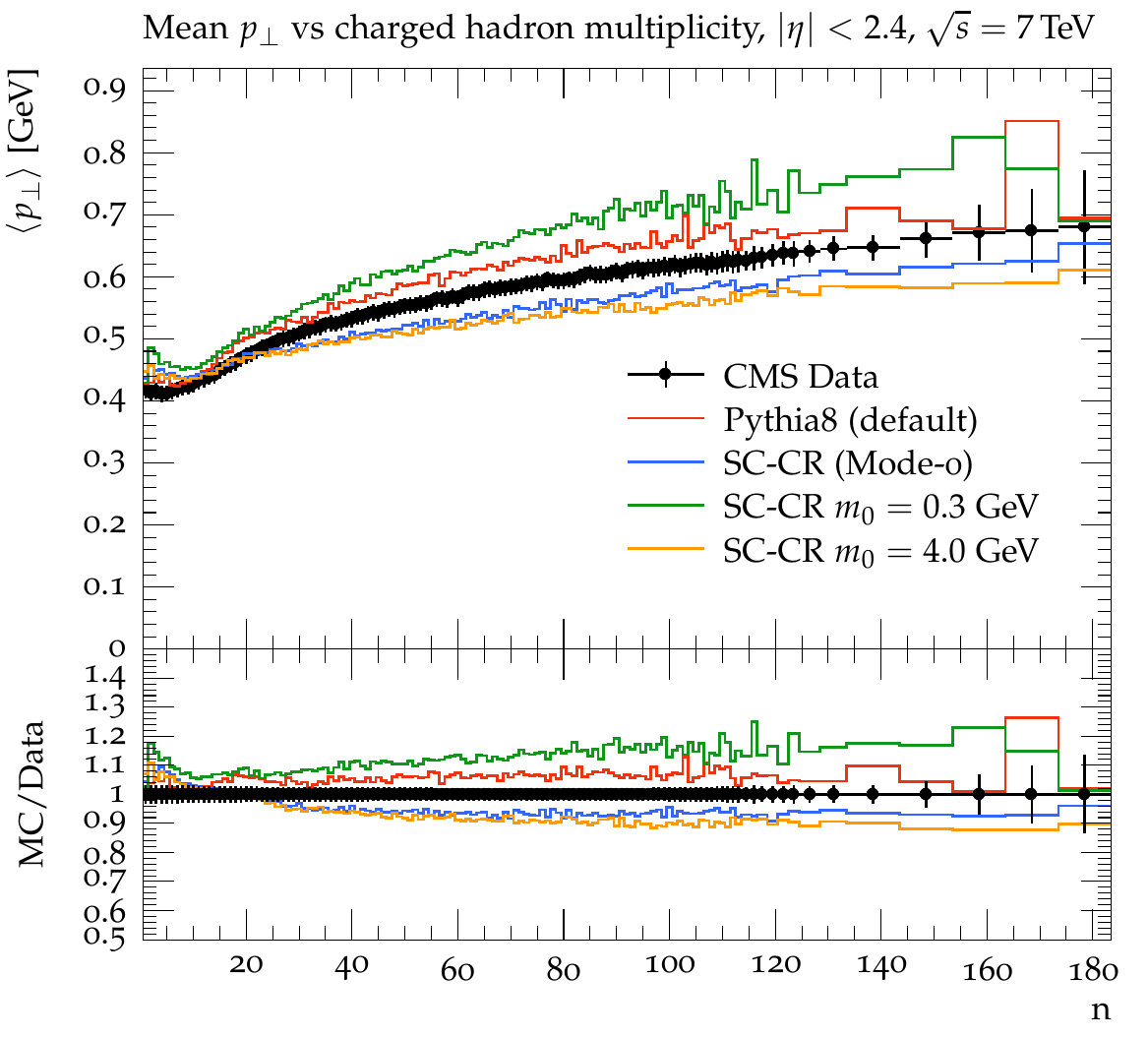}
\end{subfigure}
\caption{The same as Figure \ref{fig:pp_ptref2_mult}, but here the green and orange lines show the effect of varying the $m_0$ parameter in the SC-CR model.}
\label{fig:pp_m02_mult}
\end{figure*}

Figure \ref{fig:pp_m02_mult} shows the effect of varying $m_0$ on the
final state charged multiplicity and
$\langle p_\perp \rangle (N_{ch})$ compared with CMS data for
$\sqrt{s}=7$~TeV \pp NSD collisions. It is evident that increasing
$m_0$ increases the event multiplicity, by reducing the number of
dipoles in CR, and vice versa.

\begin{figure*}
\begin{subfigure}{.5\textwidth}
  \includegraphics[width=\linewidth]{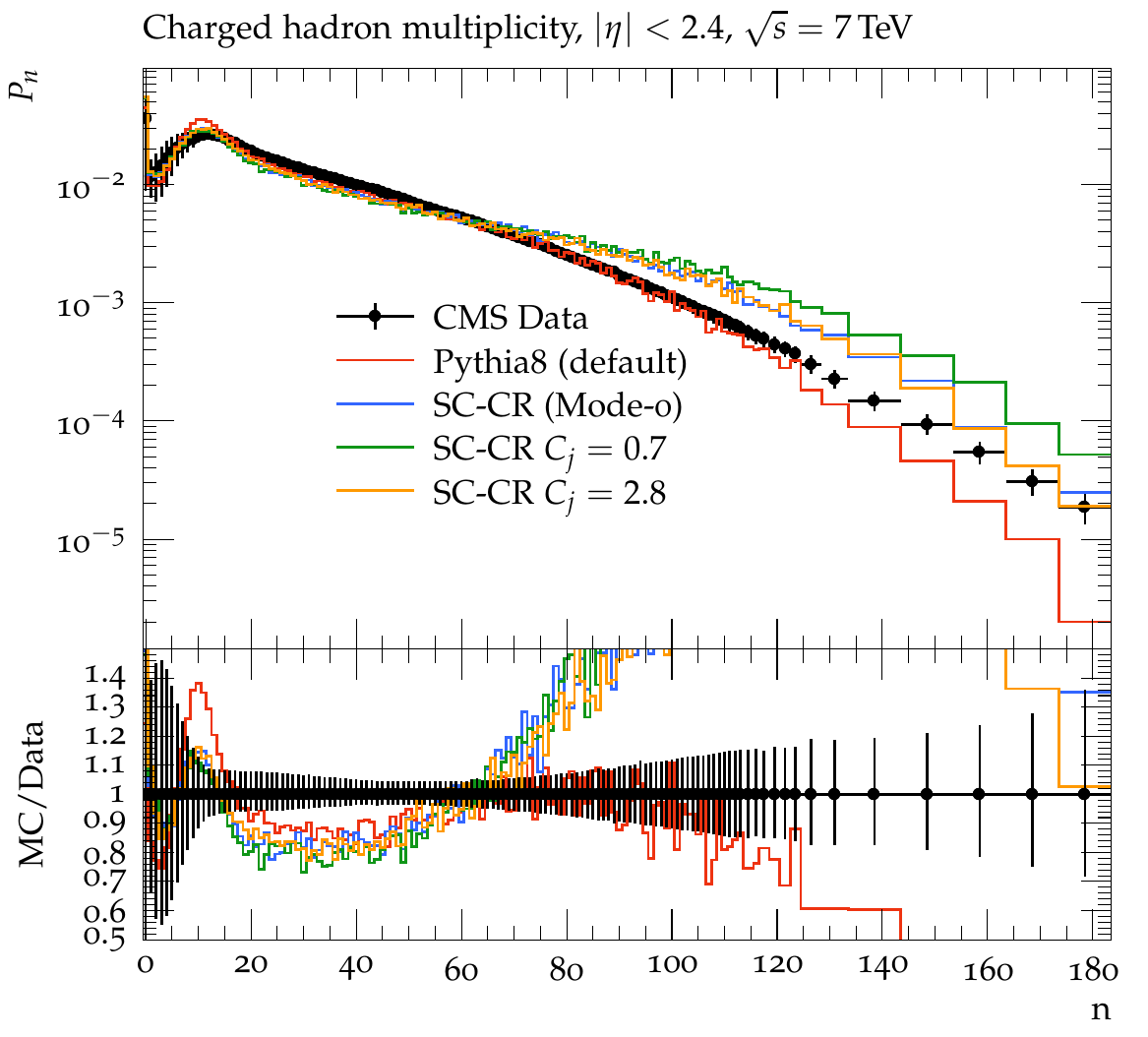}
\end{subfigure}%
\begin{subfigure}{.5\textwidth}
  \includegraphics[width=\linewidth]{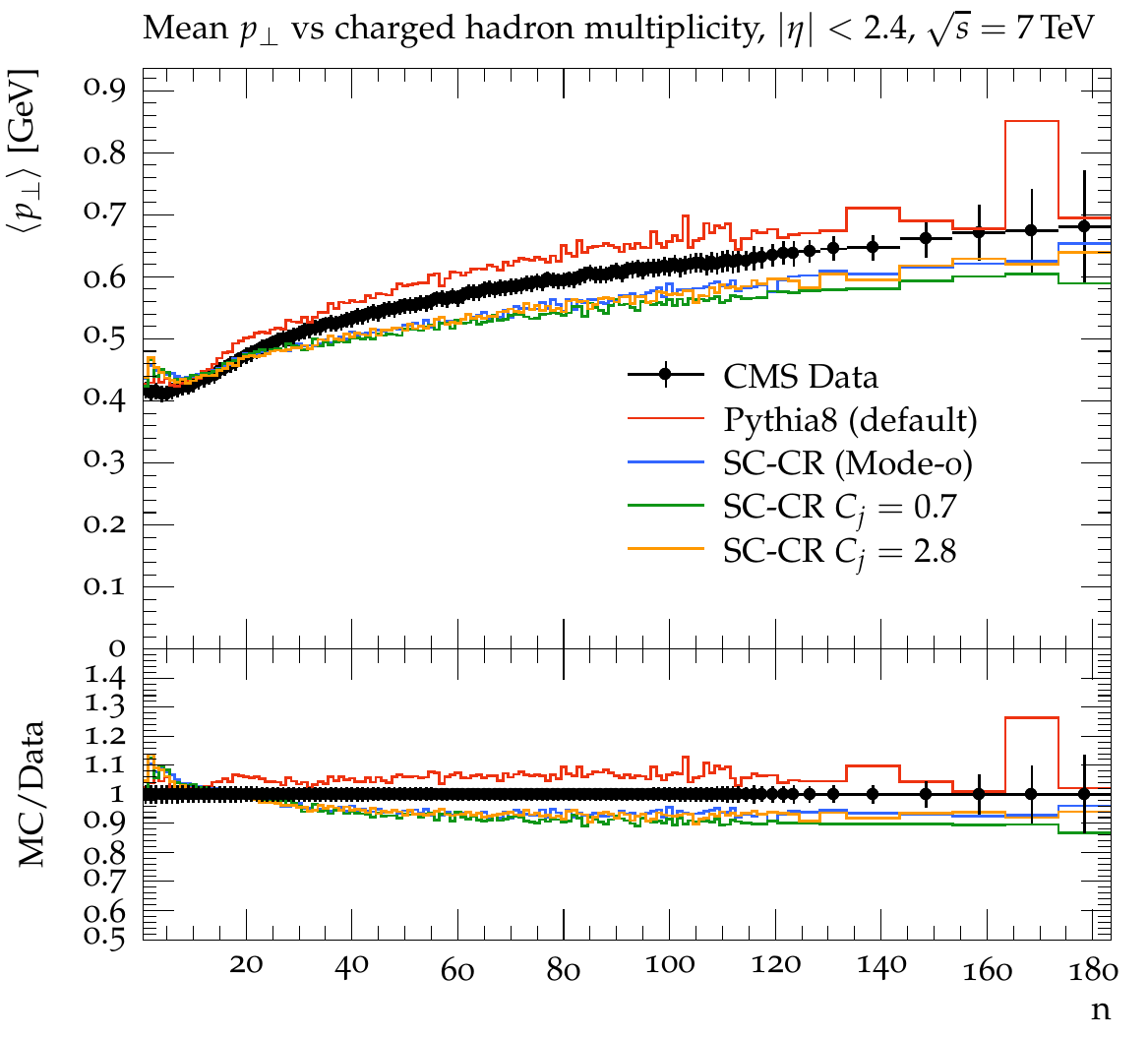}
\end{subfigure}
\caption{The same as Figure \ref{fig:pp_ptref2_mult}, but here the
  green and orange lines show the effect of varying the $C_j$
  parameter in the SC-CR model.}
\label{fig:pp_cj2_mult}
\end{figure*}

Figure \ref{fig:pp_cj2_mult} shows the effect of varying $C_j$ on the
final state charged multiplicity and
$\langle p_\perp \rangle (N_{ch})$ compared with CMS data for
$\sqrt{s}=7$~TeV \pp NSD collisions. The histograms show that reducing
$C_j$ below 1 increases the multiplicity because it reduces the number of
junctions, which allows the production of many light hadrons. But
enhancing its value will not have any significant effect on the
observables. It is also evident from the histograms that varying $m_0$
has relatively strong effects on the observables compared to varying
$C_j$.

\subsection{Allowed dipole separation}

\begin{figure*}[t]
\begin{subfigure}{.5\textwidth}
  \includegraphics[width=\linewidth]{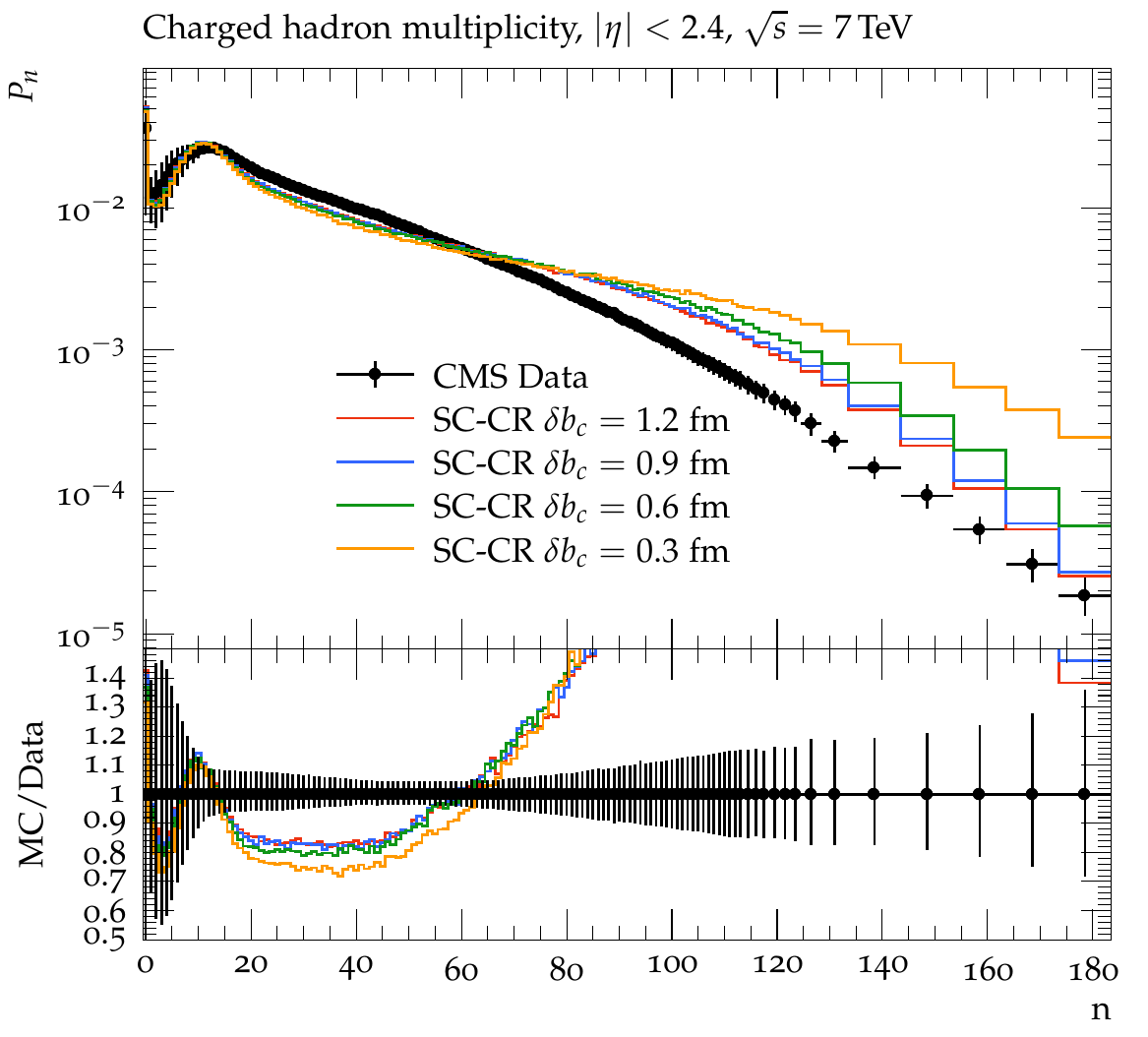}
\end{subfigure}%
\begin{subfigure}{.5\textwidth}
  \includegraphics[width=\linewidth]{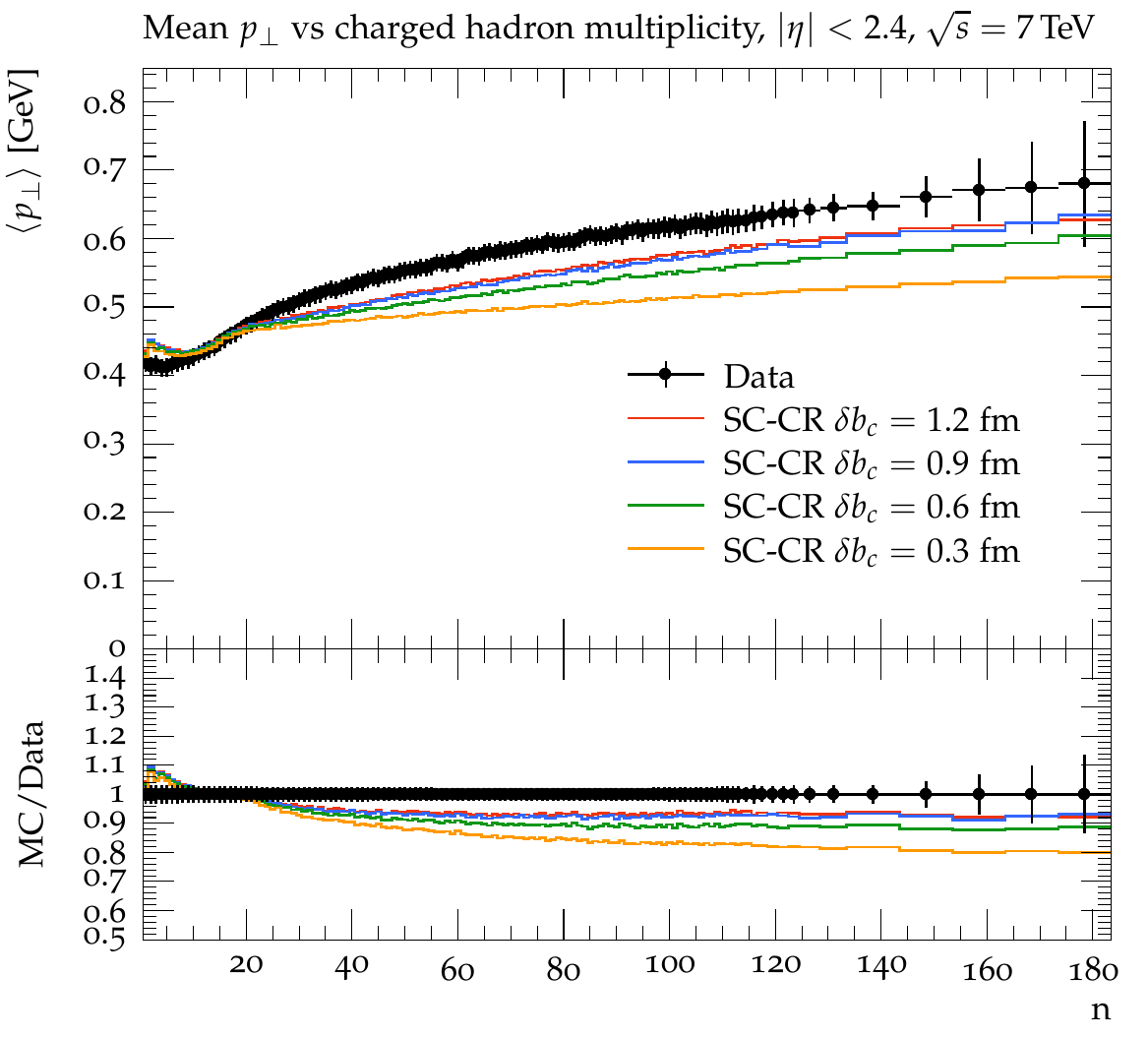}
\end{subfigure}
\caption{Comparing variation in the \ADS parameter in \pythia. Events
  are generated for $\sqrt{s}=7$~TeV \pp NSD collisions and compared
  with CMS data \cite{CMS:2010qvf}. $\textit{Left}$: charged hadron multiplicity
  in $\mid \eta \mid$ < 2.4. $\textit{Right}$: $\langle p_\perp \rangle$ vs
  $N_{ch}$ for $\mid \eta \mid$ < 2.4.}
\label{fig:pp_ADS2_mult}
\end{figure*}

The parameter \ADS constrains colour reconnection between the colour dipoles by
constraining the transverse separation. We now fix all the parameters
to QCD-CR (mode-0) and vary the \ADS. The effect on the event
multiplicity and $\langle p_\perp \rangle (N_{ch})$ due to varying
\ADS between two and three dipoles to be colour reconnected is shown
in Figure \ref{fig:pp_ADS2_mult}, and it is compared with CMS data
for $\sqrt{s}=7$~TeV \pp NSD collisions. From the figure, we see that
reducing \ADS as low as 0.3~fm will reduce the CR
significantly, increase the charged multiplicity and make the $\langle p_\perp \rangle (N_{ch})$ distribution flatter.

In a nutshell, each of the parameters discussed above has their
contribution to the total charged multiplicity, which is summarised in
table \ref{tab:parameters}. The direction of the arrows in the
bracket next to every parameter shows the direction in which the
histogram lines will move with respect to the earlier value of the
parameter if all the other parameters are fixed.

\begin{table}[!htb]
\centering
\begin{tabular}{l*{1}{c}}
Parameters & Charged Multiplicity \\
\hline
$p^{\text{ref}}_{\perp0}$ ($\downarrow$) & $\uparrow$ \\
\hline
$m_0$	($\downarrow$) & $\downarrow$ \\
\hline
$C_j$ ($\downarrow$) & $\uparrow$ \\
\hline
\ADS ($\downarrow$) & $\uparrow$ \\
\end{tabular}
\caption{A list of parameters and the effects on the overall
  multiplicity in \pythia. When the parameter is reduced
  $(\downarrow)$ the overall multiplicity will increase $(\uparrow)$
  or decrease ($\downarrow$), when keeping all other parameters fixed.}
\label{tab:parameters}
\end{table}

\subsection{$\epsilon_{pom}$ for SND events}

The default Angantyr simulates SND events using a proton-like pomeron
PDF, a pomeron-proton interaction cross-section similar to the
proton-proton non-diffractive interaction cross-section, and the pomeron
flux according to Schuler and Sj{\"o}strand \cite{Schuler:1993wr},
which gives a logarithmic distribution in the mass of the diffracted
system, $dm^2/m^2$.

\begin{figure*}
\centering
\begin{subfigure}{.5\textwidth}
  \includegraphics[width=\linewidth]{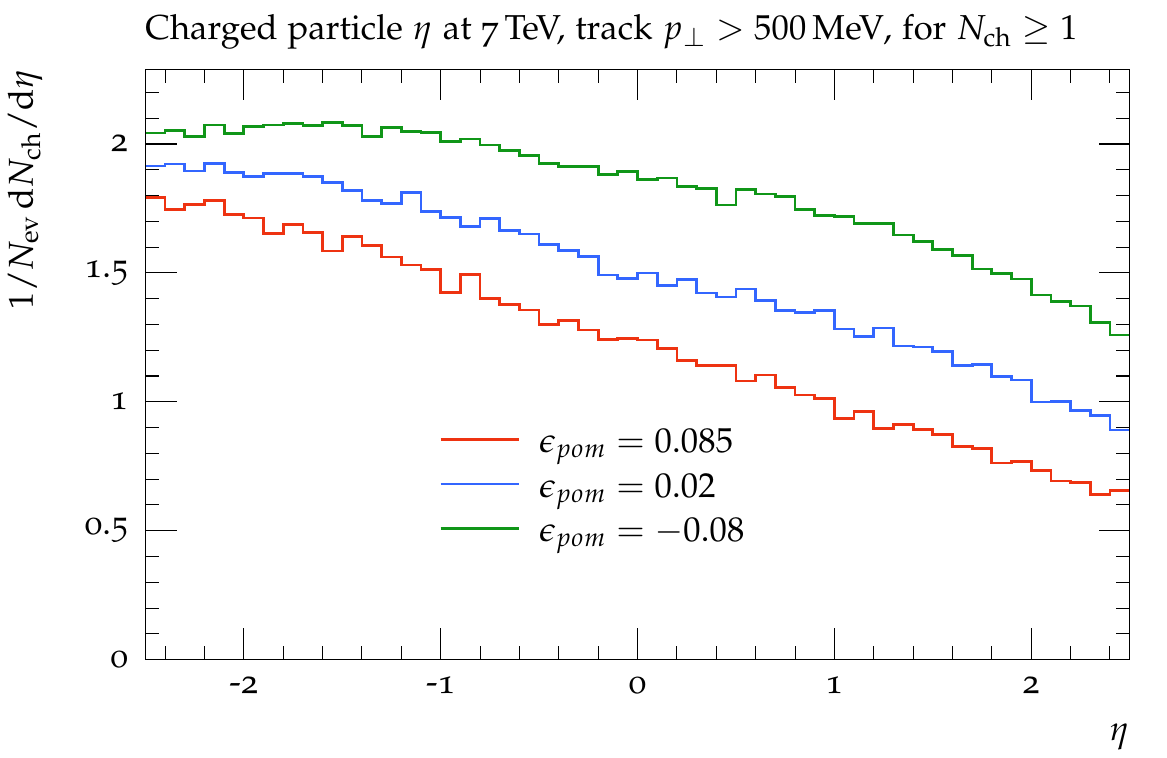}
  \caption{$\eta$ distribution for $N_{ch} > 1$}
\end{subfigure}%
\begin{subfigure}{.5\textwidth}
  \includegraphics[width=\linewidth]{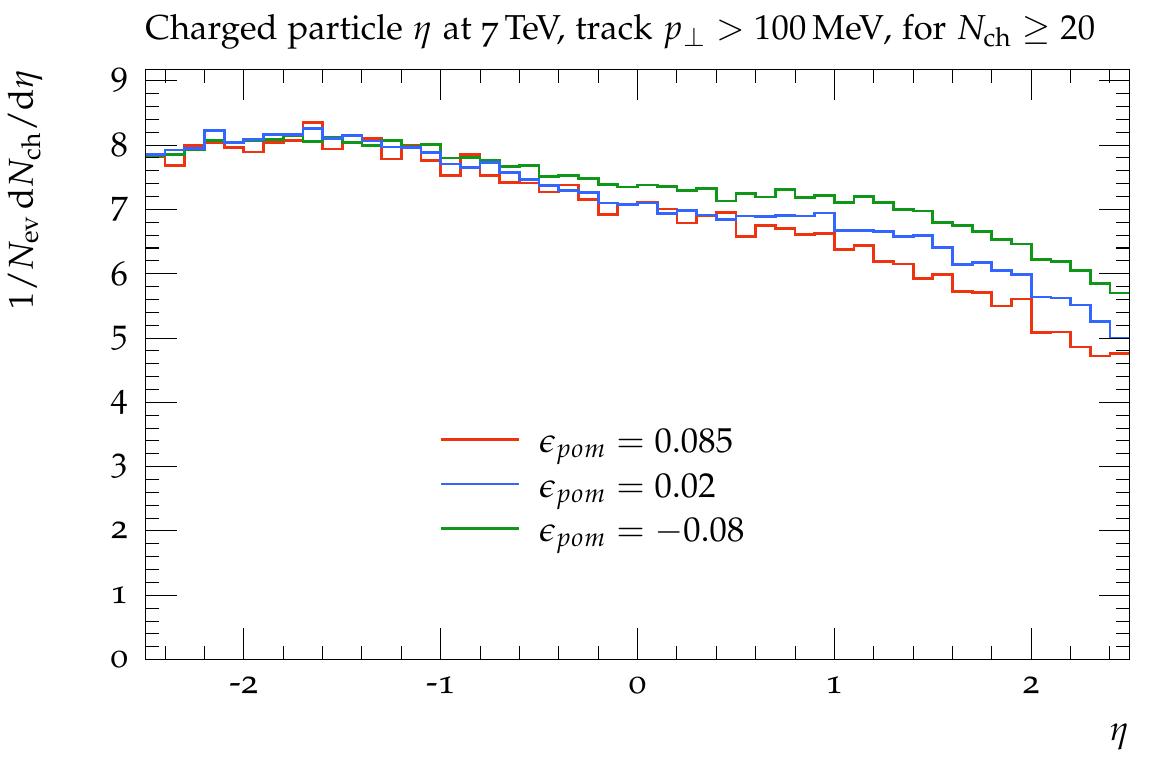}
  \caption{$\eta$ distribution for $N_{ch} > 20$}
\end{subfigure}
\caption{Angantyr SND events, which are single diffractive (SD) events
  generated for $\sqrt{s}=7$~TeV \pp collisions, with
  \setting{HeavyIon:Mode = 2}, and \setting{Angantyr:SDTests = on} in
  \pythia. Here the SD events are generated in the
  \setting{Angantyr:SASDmode = 4}. Changing $\epsilon_{pom}$ from
  positive to negative values, the event multiplicity increases, and
  the overall distribution is a bit flatter, but closer to the
  diffractive proton side (negative $\eta$), its similar to ND as
  expected.}
\label{fig:epom1_atlas_mult}
\end{figure*}

When we now introduce CR also between different sub collisions in a HI
event we expect the overall multiplicity to go down, possibly
destroying the good reproduction of data reported in
\cite{Angantyr}. To compensate for this we want to modify the pomeron
flux in the SND events, and we use a conventional supercritical description for the pomeron
flux, attributed to Berger $\textit{et al.}$ \cite{Berger:1986iu} and
Streng \cite{Streng:1988hd}.  The Pomeron Regge trajectory is
parameterized as:
\begin{equation}
  \alpha(t) = 1 + \epsilon_{pom} + \alpha^{'}(t ),
\end{equation}
giving the mass distribution $dm^2/m^{2(1+\alpha(t))}$.  We can then
vary $\epsilon_{pom}$, which is a parameter to modify the mass
distribution in the diffracted system.

The effect of changing the $\epsilon_{pom}$ parameter for the low and
high multiplicity SND events in Angantyr is shown in Figure
\ref{fig:epom1_atlas_mult}.  As expected we see an increase in the
multiplicity when changing $\epsilon_{pom}$ from positive to negative
values.

It should be noted that the default value of $\epsilon_{pom}$ in
\pythia\ is 0.085, which gives a good fit for diffractive events in
\pp. However, when we here use this to generate SND events, we do not
expect them to behave exactly like diffractive events. In
\cite{Angantyr} we discussed the relationship between the SND in
double nucleon scattering and single diffraction (see the discussion
of Figure 22) and argued that the mass distribution should not be the
same in the two cases. The mass distribution in single diffraction is
related to the rapidity span of the diffracted system,
$dm^2/m^2\approx d\Delta y_m$, while in the SND we expect it to be
proportional to the rapidity gap, $dm^2/m^2\approx d\Delta y_{gap}$,
single diffraction, with $\Delta y_{gap}=\Delta Y - \Delta y_m$, where
$\Delta Y$ is total rapidity span of the \pp\ collision. Therefore, if
a positive $\epsilon_{pom}$ is needed to describe single diffraction,
using a negative value is quite reasonable for the SND, which is what
we need in order to compensate for the decrease in multiplicity due to the
CR in HI collision.

\subsection{CR effects in \pA and \AA collisions}

\begin{figure*}
\centering
\begin{subfigure}{.5\textwidth}
  \includegraphics[width=\linewidth]{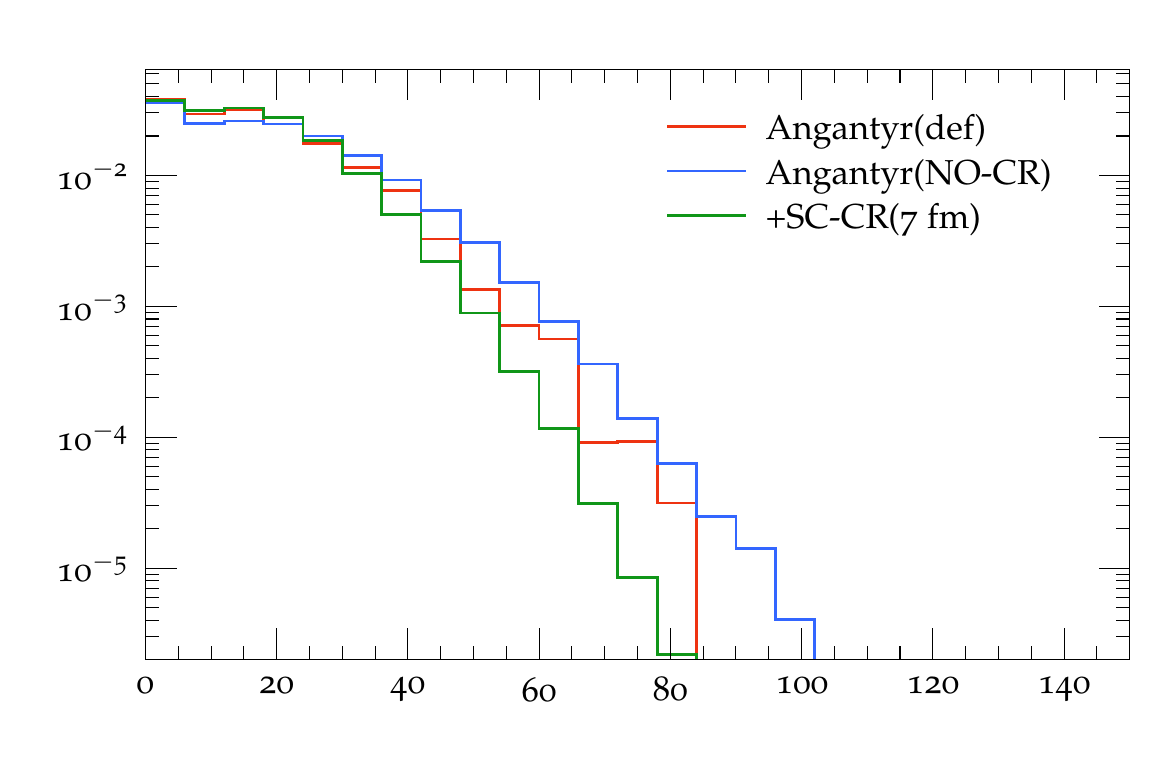}
\end{subfigure}%
\begin{subfigure}{.5\textwidth}
  \includegraphics[width=\linewidth]{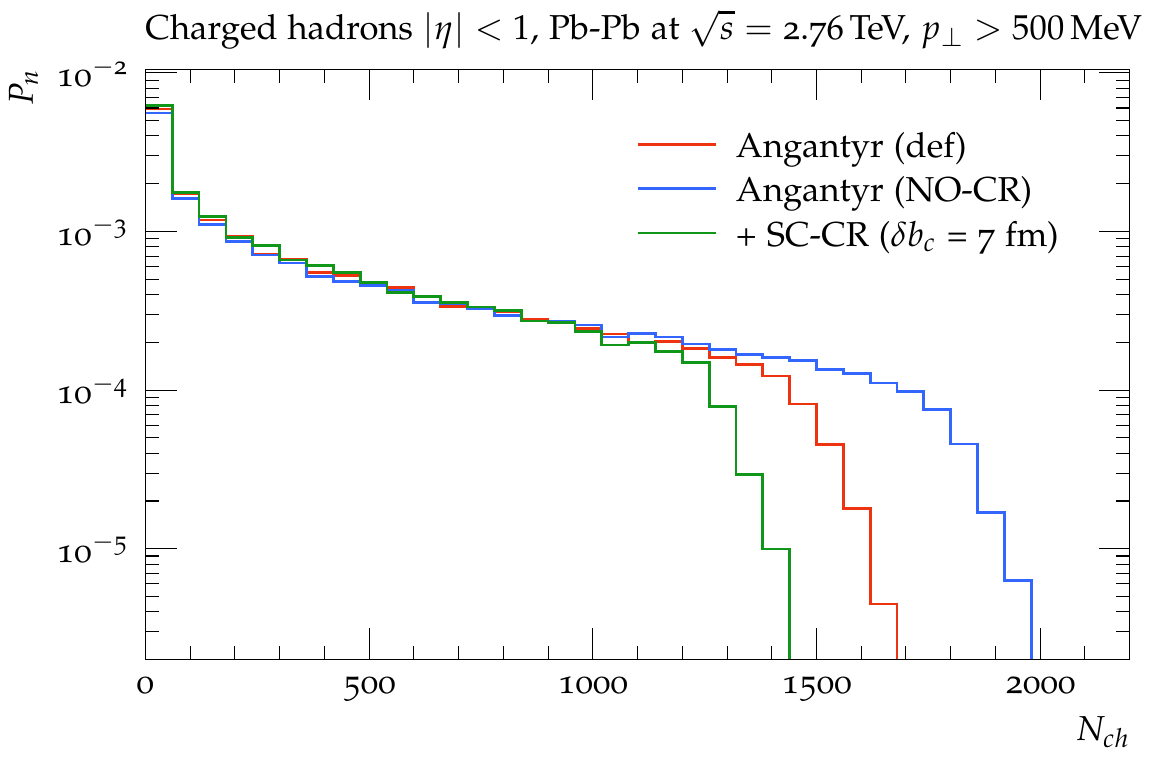}
\end{subfigure}%
\caption{The distribution in charged hadrons multiplicity in
  $\mid \eta \mid$ < 1 and with $p_{\perp} > 500 \mathrm{MeV}$ for
  \pPb events at $\sqrt{s_{NN}} = 5.02$~TeV (left) \PbPb events at
  $\sqrt{s_{NN}} = 2.76$~TeV (right). The red lines are for the events
  generated with the default Angantyr. The blue lines are the events
  with no CR reconnection between the colour dipoles, and the green
  lines are the events generated with QCD-CR (mode-0) parameters, but
  the $\ADS$ value is set to 7 $\mathrm{fm}$, which is almost the
  radius of a Pb nucleus.}
\label{fig:cr_checks}
\end{figure*}

This is the first time that the effects of CR have been introduced and
studied in a heavy-ion collision event-generator.  In section
\ref{S:cr}, we show the importance of the CR in the context of \ee and
\pp collision event simulations in \pyt.  In this work, we are further
extending the Angantyr model of \pythia with a global CR, which is
constrained by the transverse separation of the colour dipoles.  For
the sake of completeness, it is interesting to see how large the
effects of colour reconnections really are before retuning. We have
looked at \pPb and \PbPb\ central charged event multiplicities for the
default Angantyr setup, where there are colour reconnections only
inside individual \NN\ sub-collisions. This we then compared to the
case where reconnections are switched off altogether, and to the case
where we have global reconnections between (almost) all dipoles in the
event. We have used the QCD-CR (mode-0) parameters for the latter, but
the $\ADS$ value is set to 7 $\mathrm{fm}$.

The results are shown in Figure \ref{fig:cr_checks}, where it is clear
that the effects of colour reconnections are substantial mainly for the
highest multiplicities. The average multiplicities are however only
moderately affected, with a 20\% increase for \PbPb when reconnections
are switched off, and a 10\% decrease with global reconnections. The
corresponding effects in \pPb are smaller, with $+10$\% and $-5\%$
respectively.

\section{Results}
\label{S:results}

First, we note that we have not done a very sophisticated retuning of
the parameters, \eg, using the Professor framework
\cite{Buckley:2009bj}. Our focus has been to get a reasonable
multiplicity distribution for \pp\ and also for \pA. There is some
tension in data, favouring a larger \ADS\ in \pp, but a lower one in
\pA. In the end, we basically selected the value $\ADS=0.5$~fm.

We show the tuned values under the SC-CR (tuned) column and the
default values of those parameters under the QCD-CR (mode-0) column in
Table \ref{tab:table2}.  For \pp, all other parameters are the same as
for QCD-CR (mode-0), which is shown in Table \ref{tab:table1}.

\begin{table*}[!htb]
\centering
\begin{tabular}{l*{2}{c}}
Parameters & QCD-CR (Mode-0) & SC-CR (tuned) \\
\hline
\setting{PartonVertex:setVetex} & - &on \\
\hline
\setting{MultiPartonInteractions:pT0Ref} & 2.12 & 2.37 \\
\hline
\setting{ColourReconnection:m0} & 2.9 & 1.05 \\
\hline
\setting{ColourReconnection:junctionCorrection} & 1.43 & 1.37 \\
\hline
\setting{ColourReconnection:dipMaxDist} & - & 0.5 \\
\end{tabular}
\caption{The list of parameters and their new values compared to their
  values in the QCD-CR (Mode-0) tune. The new values are used in
  \pythia to generate \pp collisions with the spatial constraint in
  the SC-CR model.}
\label{tab:table2}
\end{table*}

To treat a heavy-ion event as a single event we want to stack all
\pp-like sub-collisions at the parton level, and then apply CR with
spatial constraints on all colour dipoles. To do so, we first turn
off colour reconnection in all individual sub-collisions
(\setting{ColourReconnection:reconnect = off}) and instead switch it on
only for the hadronization stage
(\setting{ColourReconnection:forceHadronLevelCR = on}).
\footnote{At the moment this combination works only within the
  Angantyr framework. Moreover, it does not work with \pyt's default
  MPI-based CR. For \pp collisions, this feature is not that useful,
  as there's only one proton-proton collision which is a
  single event by default. But if a user wishes, one can use the
  above combination for \pp collisions as well by setting the
  \setting{Heavyion:mode = 2} flag to generate \pp collisions within the
  Angantyr setup.
}

Now, with this feature enabled, \pA\ and \AA collision events undergo
CR only once per HI event, and after CR the entire event proceeds to
hadronization.  The list of changed and new parameters for HI
collisions is shown in Table \ref{tab:AA}.  The prefix "HI" for some
of the parameters means that they are only used for SND
sub-collisions.

\begin{table*}[!htb]
\centering
\begin{tabular}{l*{2}{c}}
Parameters & QCD-CR (Mode-0) & SC-CR (new) \\
\hline
\setting{ColourReconnection:reconnect} & on & off \\
\hline
\setting{ColourReconnection:forceHadronLevelCR} & off & on \\
\hline
\setting{MultiPartonInteractions:pT0Ref} & 2.12 & 2.37 \\
\hline
\setting{PartonVertex:setVetex} & - & on \\
\hline
\setting{ColourReconnection:m0} & 2.9 & 1.05 \\
\hline
\setting{ColourReconnection:junctionCorrection} & 1.43 & 1.37 \\
\hline
\setting{ColourReconnection:dipMaxDist} & - & 0.5 \\
\hline
\setting{HIMultiPartonInteractions:pT0Ref} & 2.12 & 2.37 \\
\hline
\setting{HISigmaDiffractive:mode} & - & 0 \\
\hline
\setting{HISigmaDiffractive:PomFlux} & - & 3 \\
\hline
\setting{HISigmaDiffractive:PomFluxEpsilon} & - & -0.04 \\
\hline
\setting{HIBeamRemnants:remantMode} & - & 1 \\
\hline
\setting{HIBeamRemnants:Saturation} & - & 5 \\
\hline
\setting{BeamRemnants:beamJunction} & off & on \\
\hline
\setting{HIBeamRemnants:beamJunction} & off & on \\
\end{tabular}
\caption{The list of parameters and their new values compared to their
  default values in the QCD-CR (Mode-0) tune. The new values are used
  in the Angantyr model to generate \pA\ and \AA collisions with the
  spatial constraint in the SC-CR model. The values for the
  parameters used to generate \pp events are not changed, and they are
  same as in Table \ref{tab:table2}, while the other parameters are same as in Table \ref{tab:table1}.}
\label{tab:AA}
\end{table*}

\subsection{\pp results}
\label{S:pp_results}

\begin{figure*}
\centering
\begin{subfigure}{.5\textwidth}
  \includegraphics[width=1\linewidth]{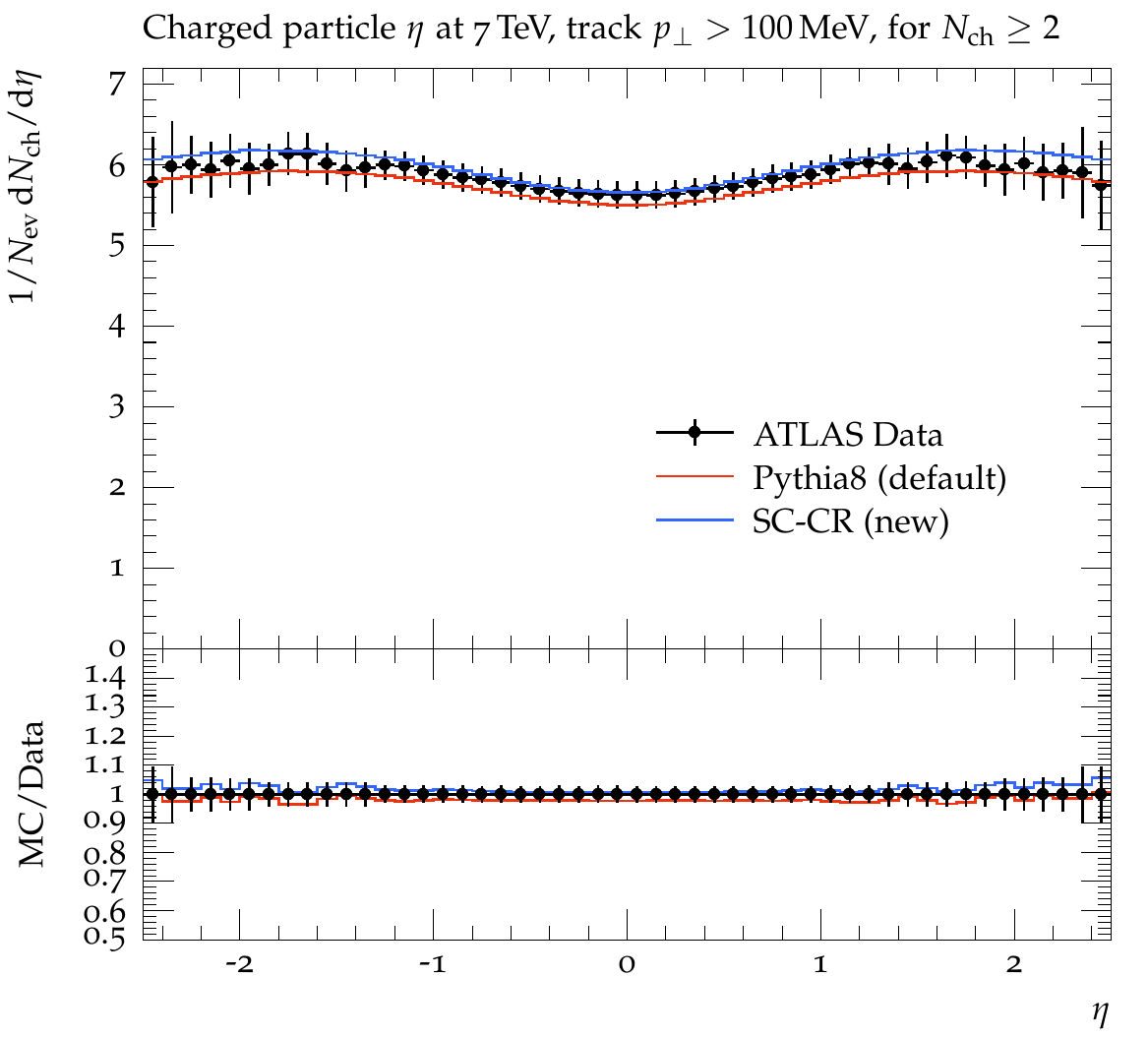}
\end{subfigure}%
\begin{subfigure}{.5\textwidth}
\includegraphics[width=1\linewidth]{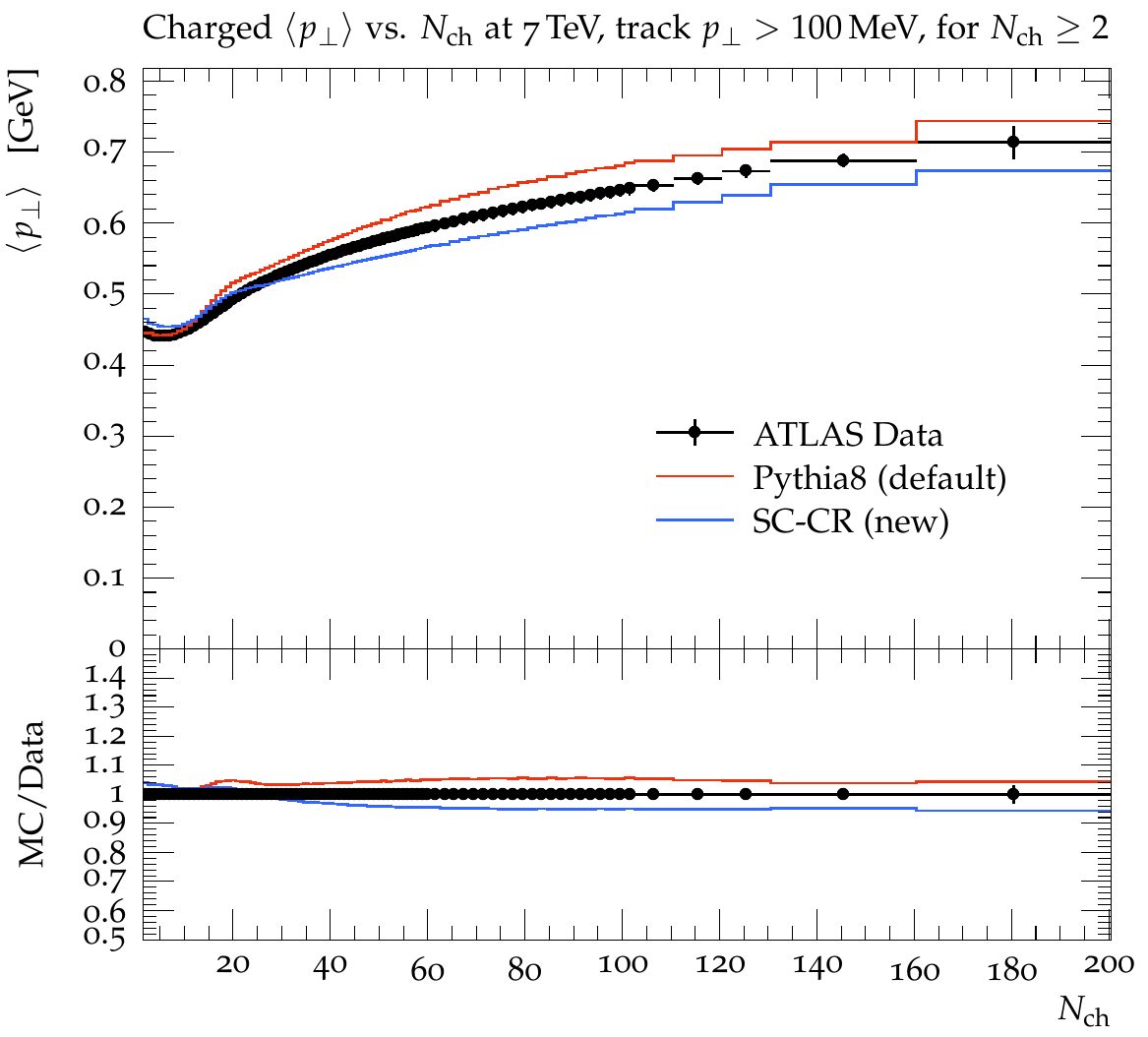}
\end{subfigure}
\begin{subfigure}{.5\textwidth}
  \includegraphics[width=1\linewidth]{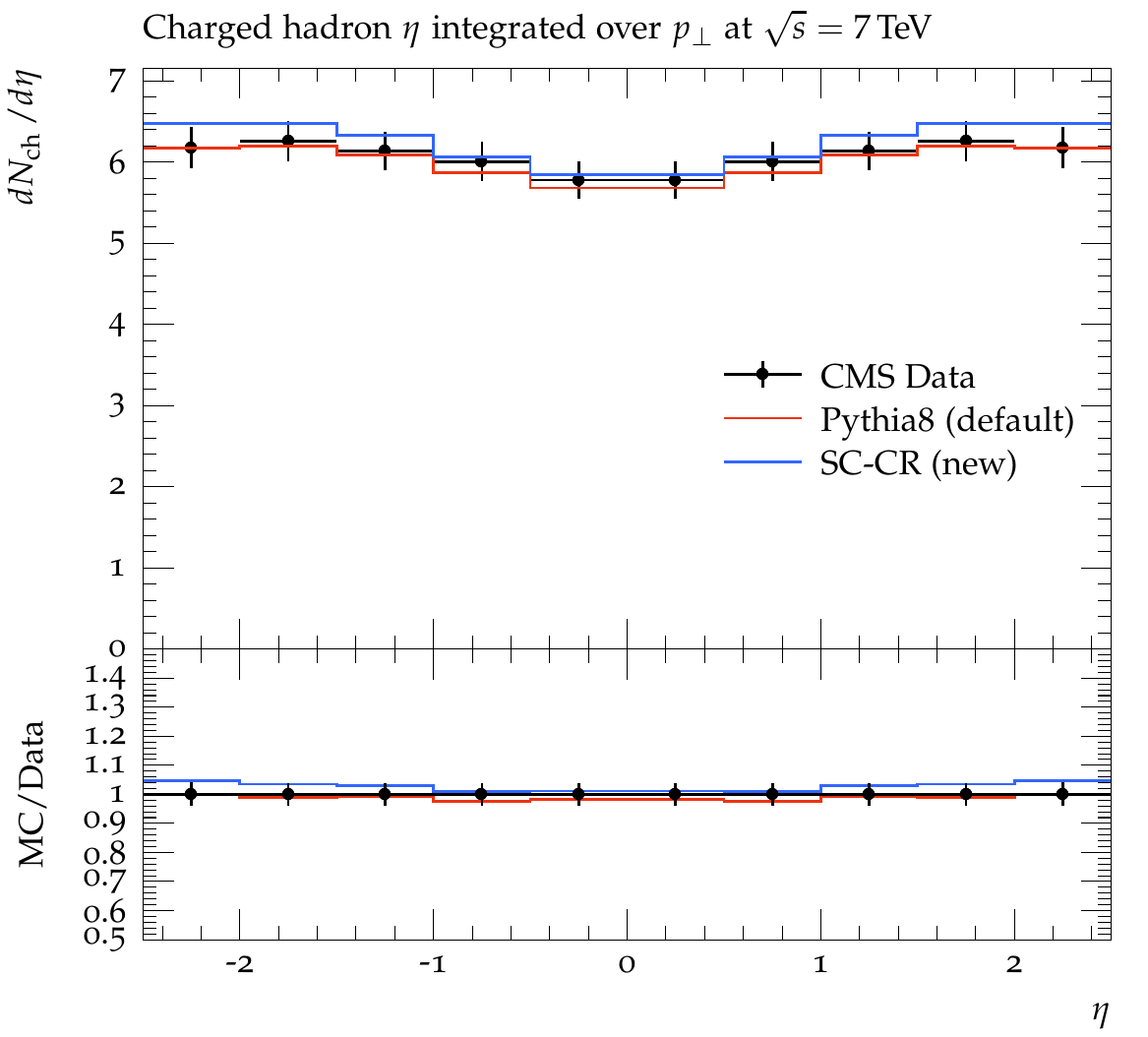}
\end{subfigure}%
\begin{subfigure}{.5\textwidth}
  \includegraphics[width=1\linewidth]{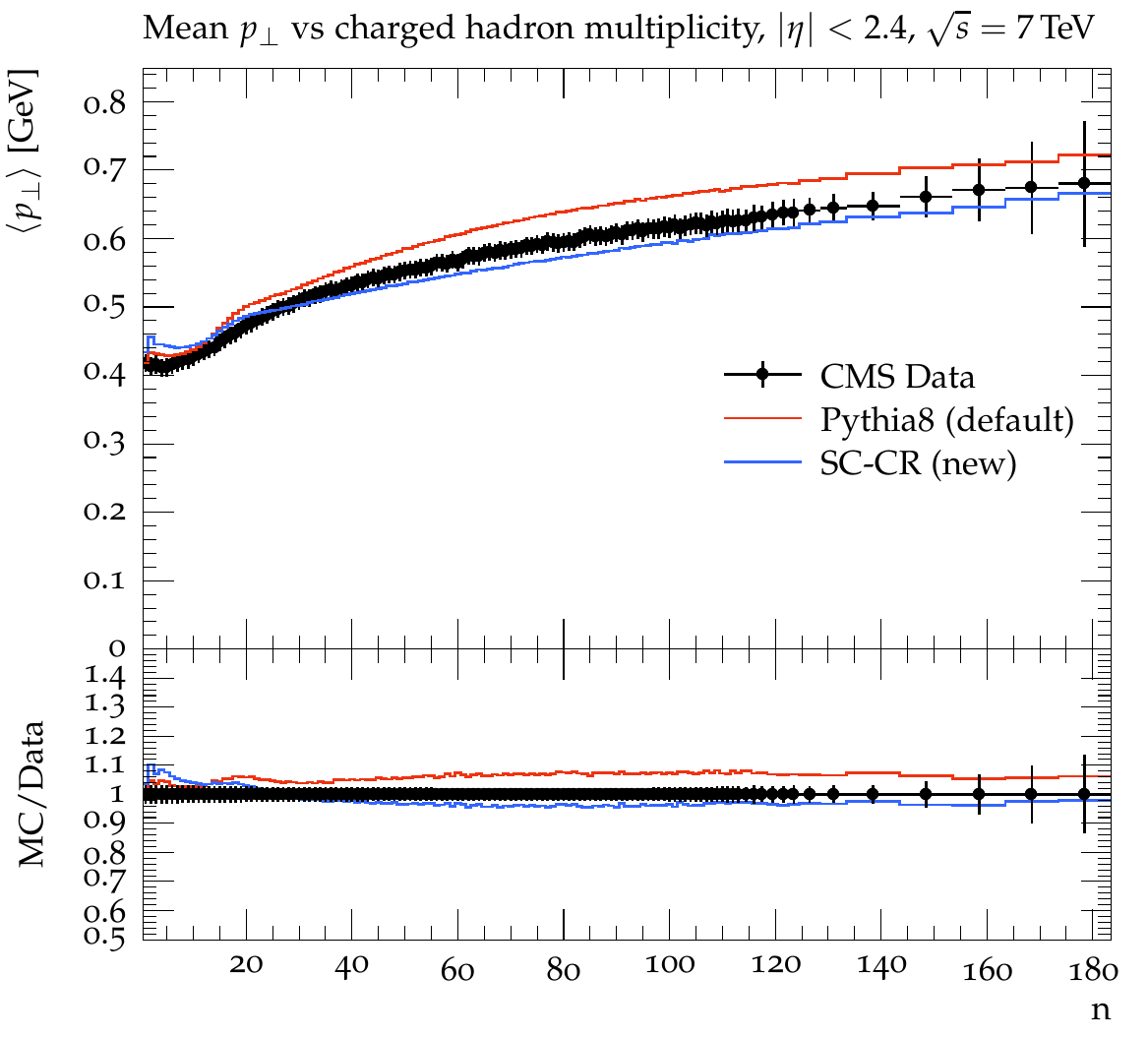}
\end{subfigure}
\caption{Events are generated for $\sqrt{s}=7$~TeV \pp
  collisions minimum bias events are compared with ATLAS
  \cite{ATLAS:2010jvh} results (top row), and
  non-single diffractive events are compared with CMS
  \cite{CMS:2010qvf,CMS:2010tjh} results (bottom row). The left plots
  show the pseudo-rapidity distribution of charged particles, while
  the right plots show $\langle p_\perp\rangle$ as a function of
  multiplicity. \pythia (default) is the default \pp collisions with
  the \pythia Monash tune setup. SC-CR (new) is produced by our spatial
  constraint CR model, with the parameters given in Table
  \ref{tab:table2}.}
\label{fig:pp_mult}
\end{figure*}

We begin with comparing \pp events generated at $\sqrt{s}=7$~TeV
using \pythia default and re-tuned spatially constrained QCD-CR in
\pythia.  We show the charged multiplicity and
$\langle p_\perp \rangle (N_{ch})$ distributions for these two setups
and compare the results with ATLAS \cite{ATLAS:2010jvh} and CMS
\cite{CMS:2010qvf,CMS:2010tjh} experiments.  The \pp results compared
with ATLAS are minimum bias events, while the events compared with the
CMS experiment are generated as non-single diffractive (NSD) events in
\pythia.  We use $c\tau > 10$ mm as the definition for primary
particles when comparing our simulated events against ATLAS and CMS
data.

\begin{figure*}
\centering
\begin{subfigure}{.5\textwidth}
  \includegraphics[width=1\linewidth]{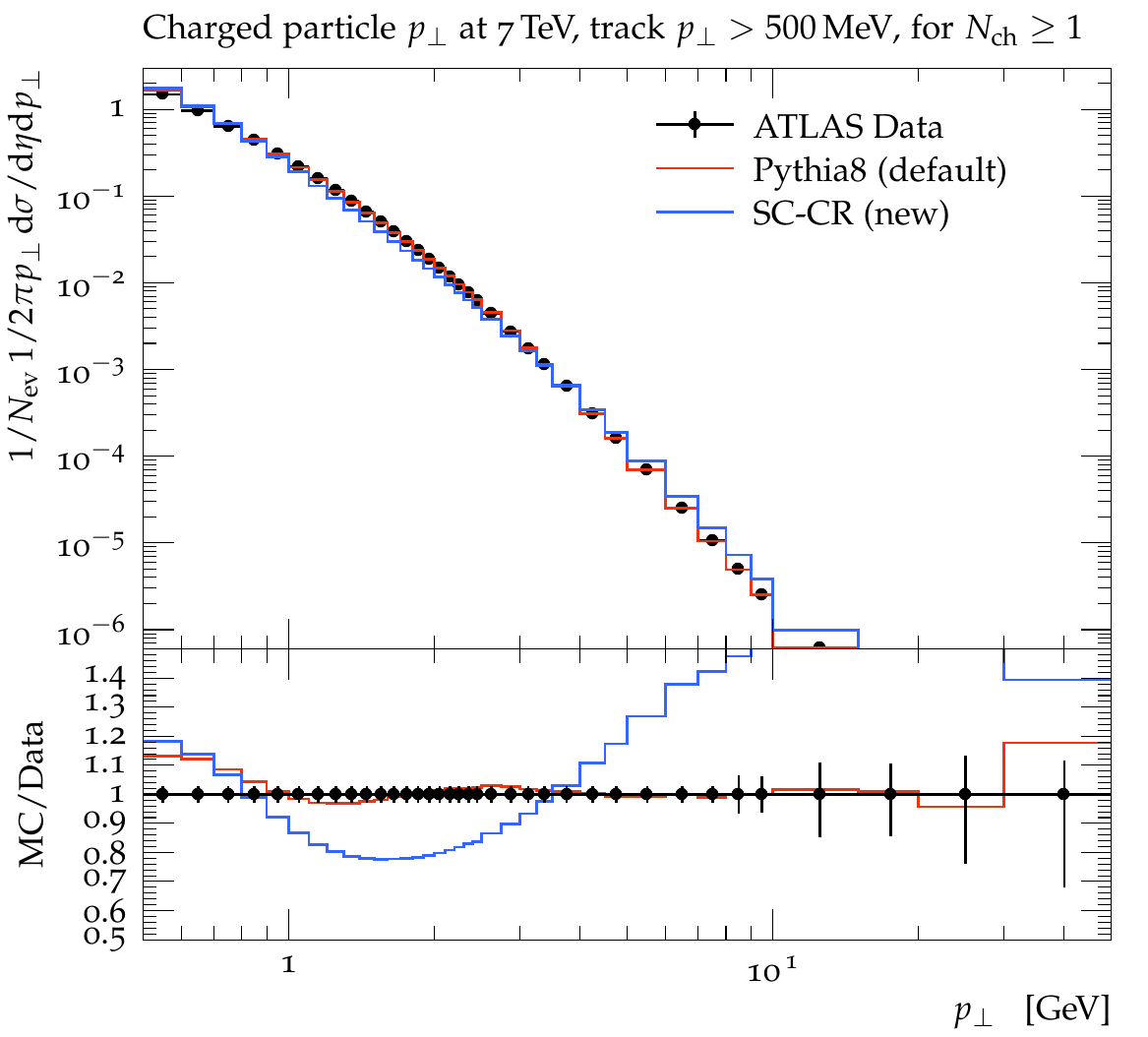}
\end{subfigure}%
\begin{subfigure}{.5\textwidth}
  \includegraphics[width=1\linewidth]{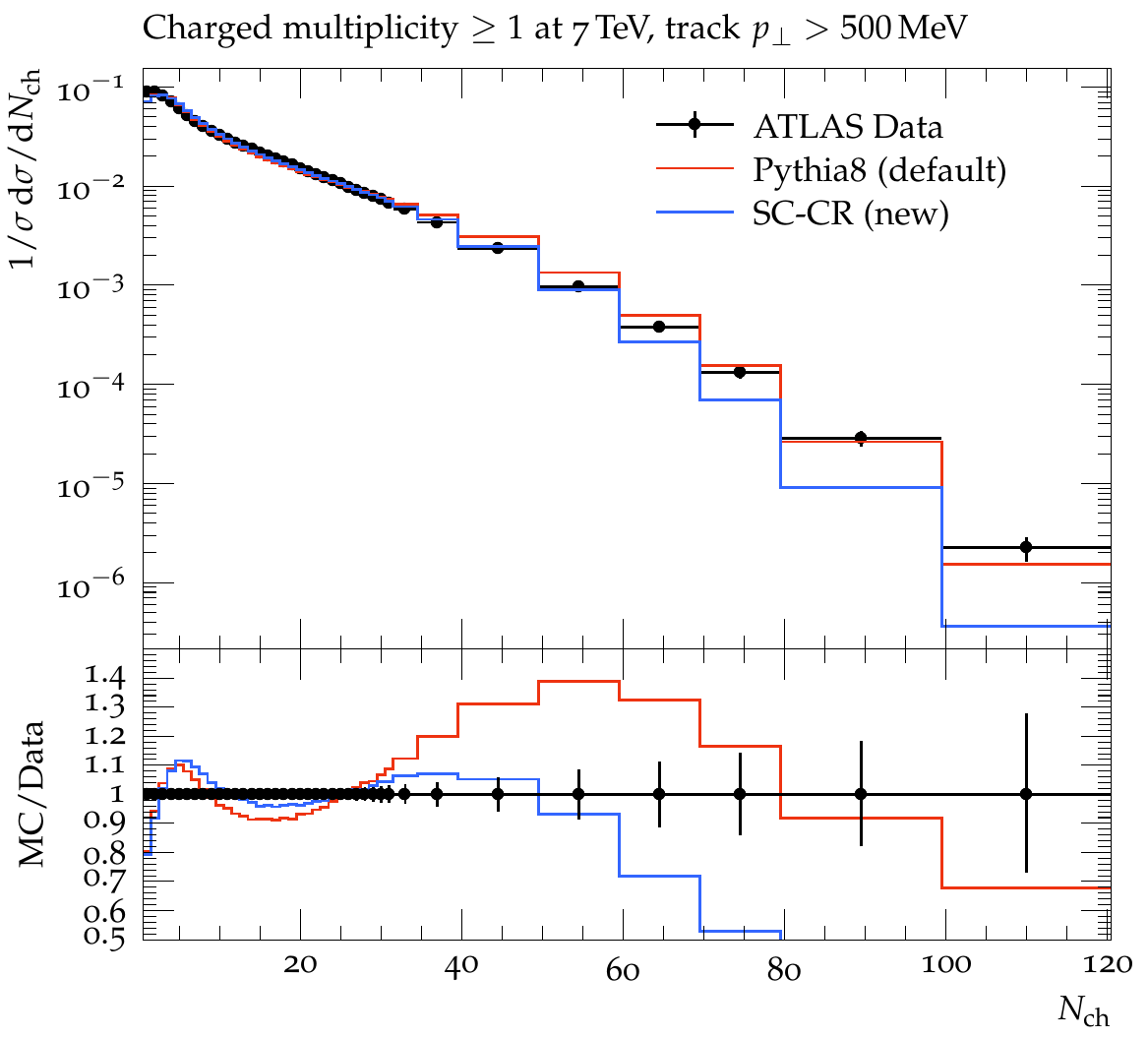}
\end{subfigure}%
\caption{Events are generated for $\sqrt{s}=$7~TeV \pp collisions
  minimum bias events and compared with ATLAS \cite{ATLAS:2010jvh}
  results. The left plot shows the distribution of $p_\perp$ for
  charged particles and the right shows the distribution in charged
  multiplicity. The lines are the same as in Figure \ref{fig:pp_mult}.}
\label{fig:pp_pt}
\end{figure*}

In figure \ref{fig:pp_mult} we show that with the new constraint on
QCD-CR and the new tuned parameters, we are able to reasonably
reproduce the average event multiplicity and
$\langle p_\perp \rangle (N_{ch})$ distributions from the ATLAS and
CMS for $\sqrt{s}=7$~TeV \pp collisions. This should not be
surprising, since these are the distributions we tuned to.

Figure \ref{fig:pp_pt} shows the distribution in pt and $N_{ch}$
compared to ATLAS data, and
here we observe that the charged multiplicity distribution for the
SC-CR setup drops too quickly for high multiplicities. We also see
that the transverse momentum spectrum becomes too hard.  These effects
are related: when we introduce the spatial constraint, there will be
less reconnection and thus higher multiplicity; this is then
compensated somewhat by decreasing $m_0$, but this is not enough
(remember that QCD-CR (mode 0) also has a too high multiplicity
when our improved junction handling is introduced); so we increase
the $p^{\text{ref}}_{\perp0}$ but that mainly decreases the
multiplicity of low transverse momentum particles, giving a too hard
spectrum. The conclusion is that our tuning may have been a bit naive,
but for now, we are satisfied with that we have the overall
multiplicity under control.

\begin{figure*}
\centering
\begin{subfigure}{.33\textwidth}
  \includegraphics[width=1\linewidth]{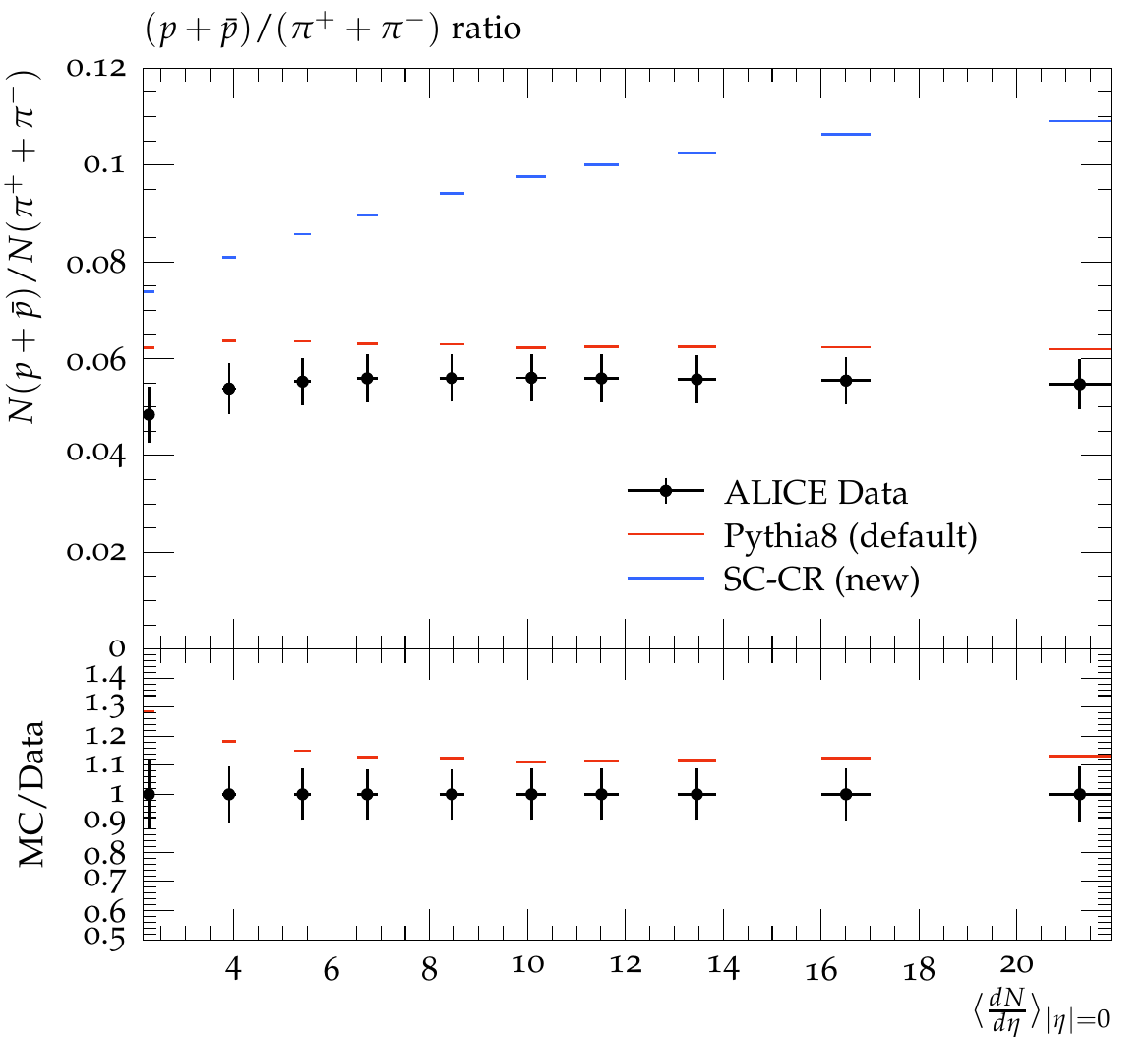}
\end{subfigure}%
\begin{subfigure}{.33\textwidth}
  \includegraphics[width=1\linewidth]{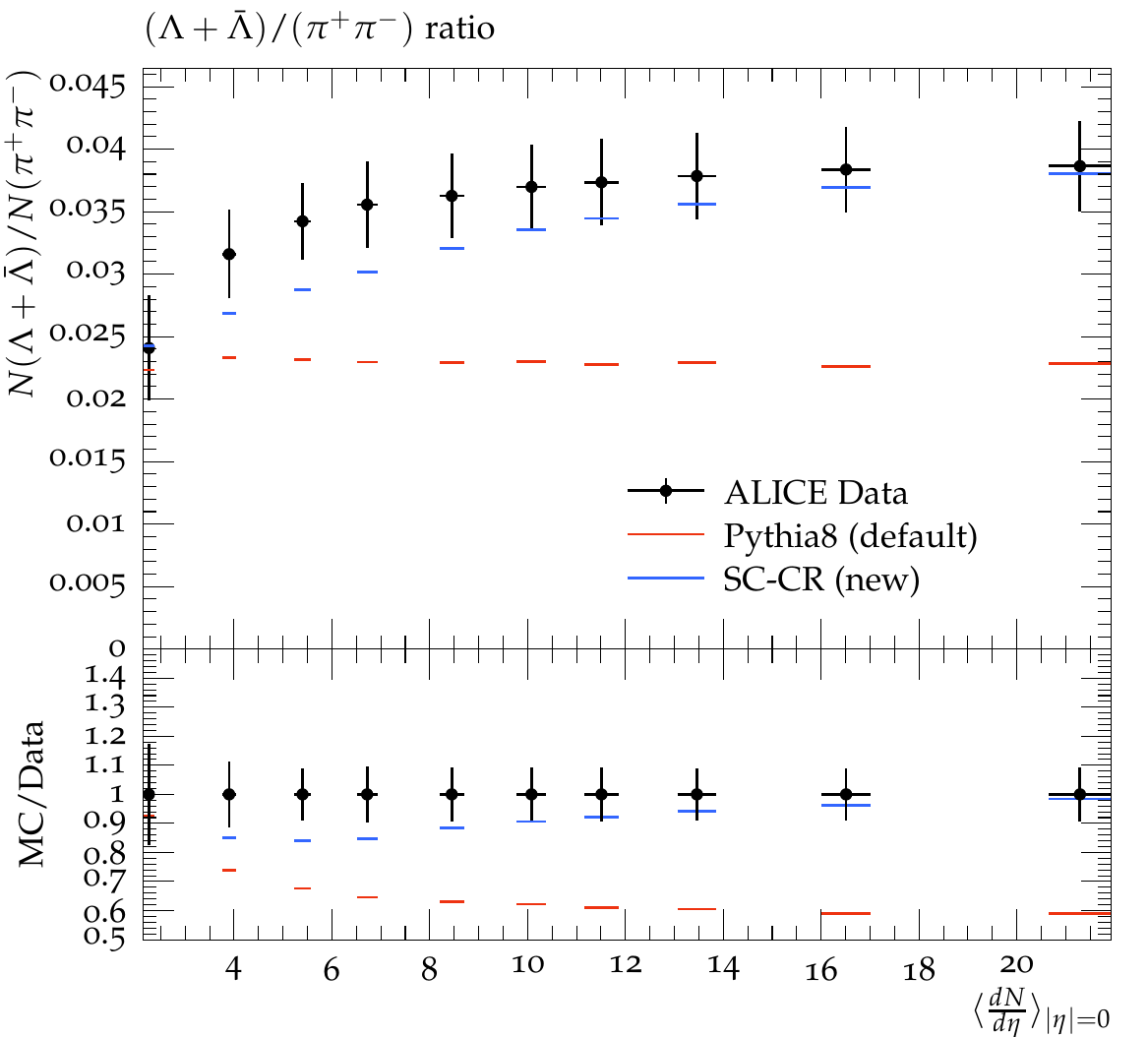}
\end{subfigure}%
\begin{subfigure}{.33\textwidth}
  \includegraphics[width=1\linewidth]{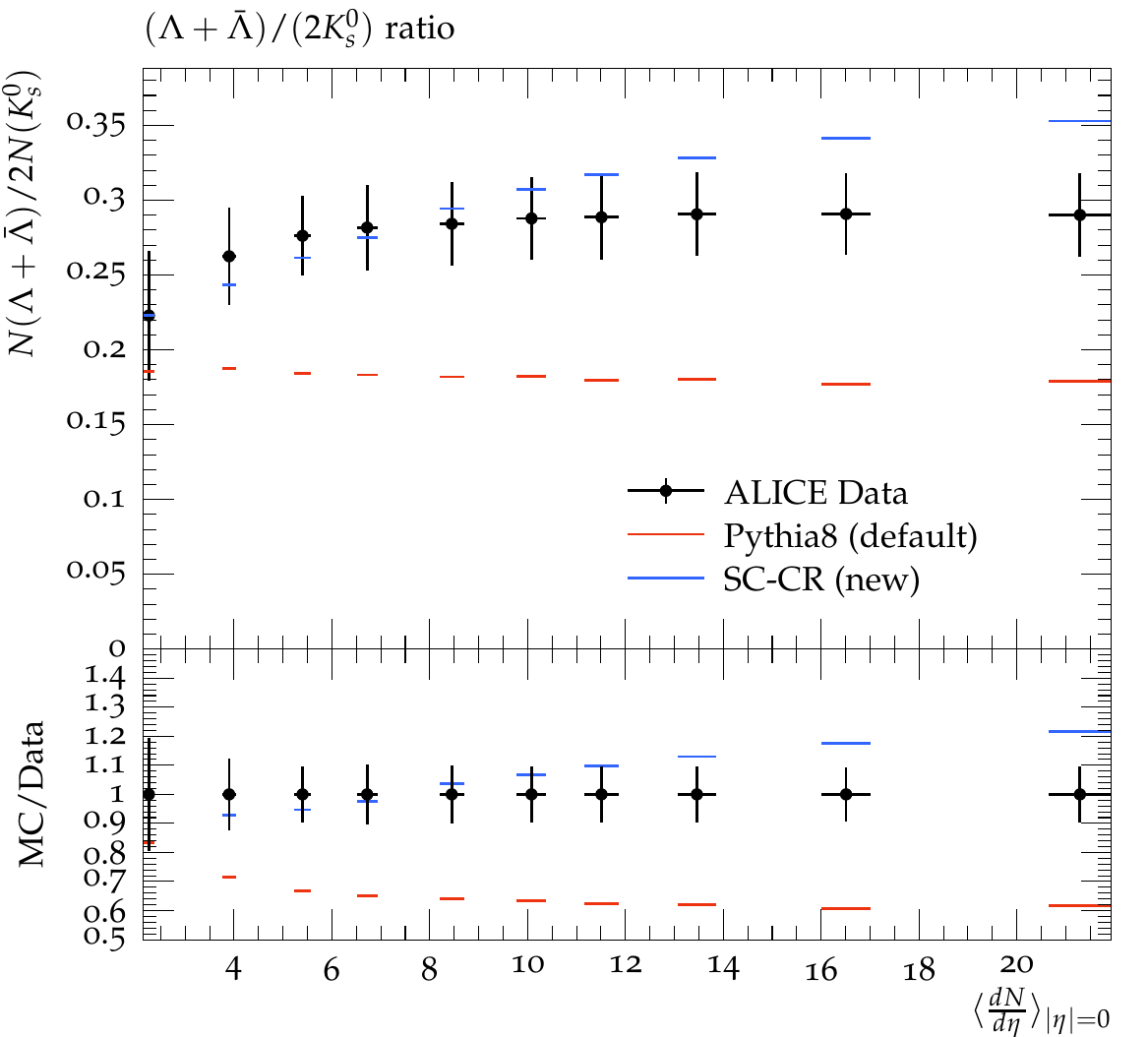}
\end{subfigure}%
\caption{From left to right the ratios of $p/\pi$, $\Lambda/\pi$, 
  and $\Lambda/K^0_s$ is plotted against the average charged-particle multiplicity in the  
  central pseudo-rapidity. \pp collisions minimum bias events are generated 
  at $\sqrt{s}=$7~TeV and compared with ALICE \cite{ALICE:2016fzo}
  results. The lines are the same as in Figure \ref{fig:pp_mult}.}
\label{fig:pp_ratio}
\end{figure*}

\begin{figure*}
\centering
\begin{subfigure}{.4\textwidth}
  \includegraphics[width=1\linewidth]{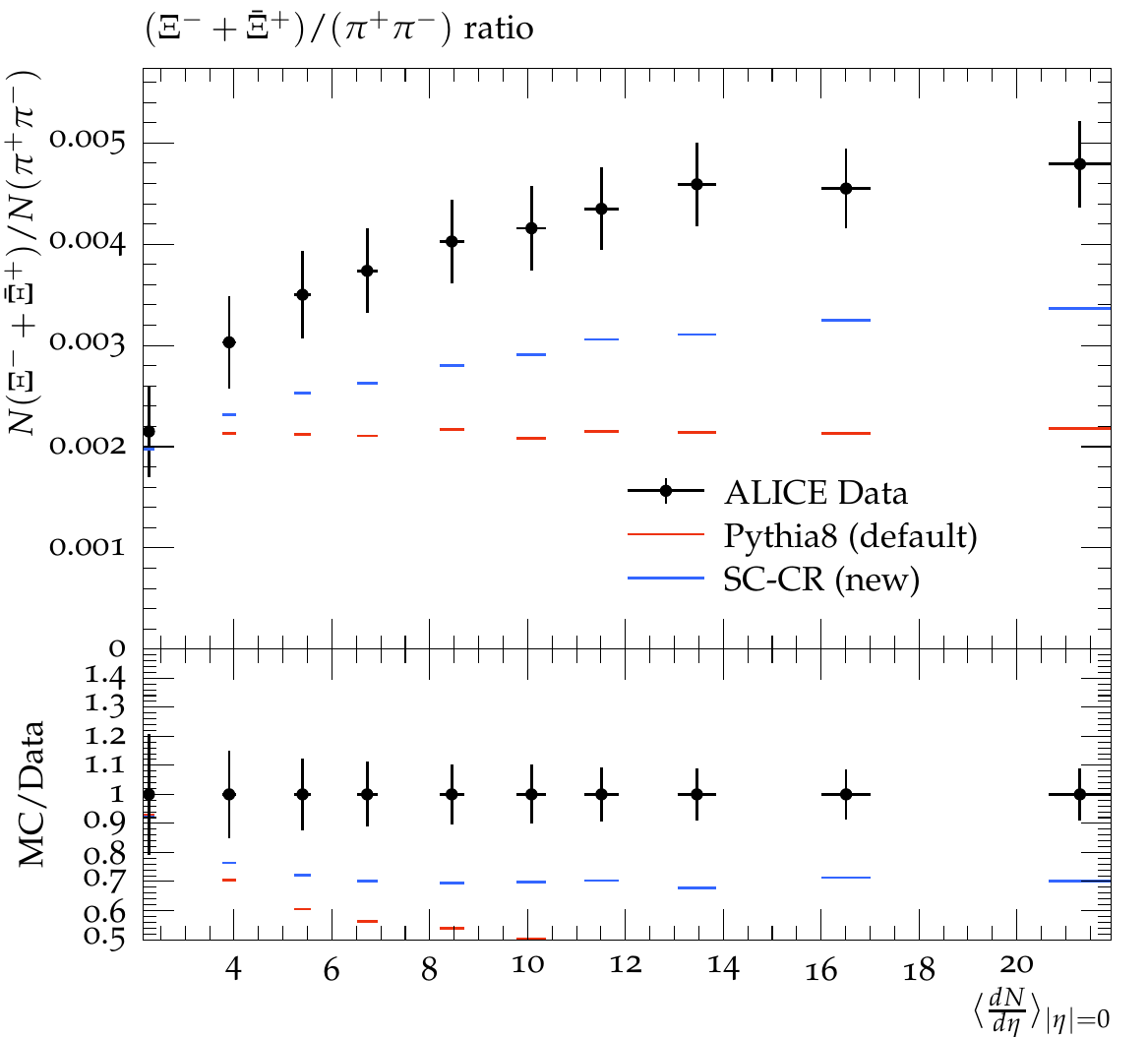}
\end{subfigure}%
\begin{subfigure}{.4\textwidth}
  \includegraphics[width=1\linewidth]{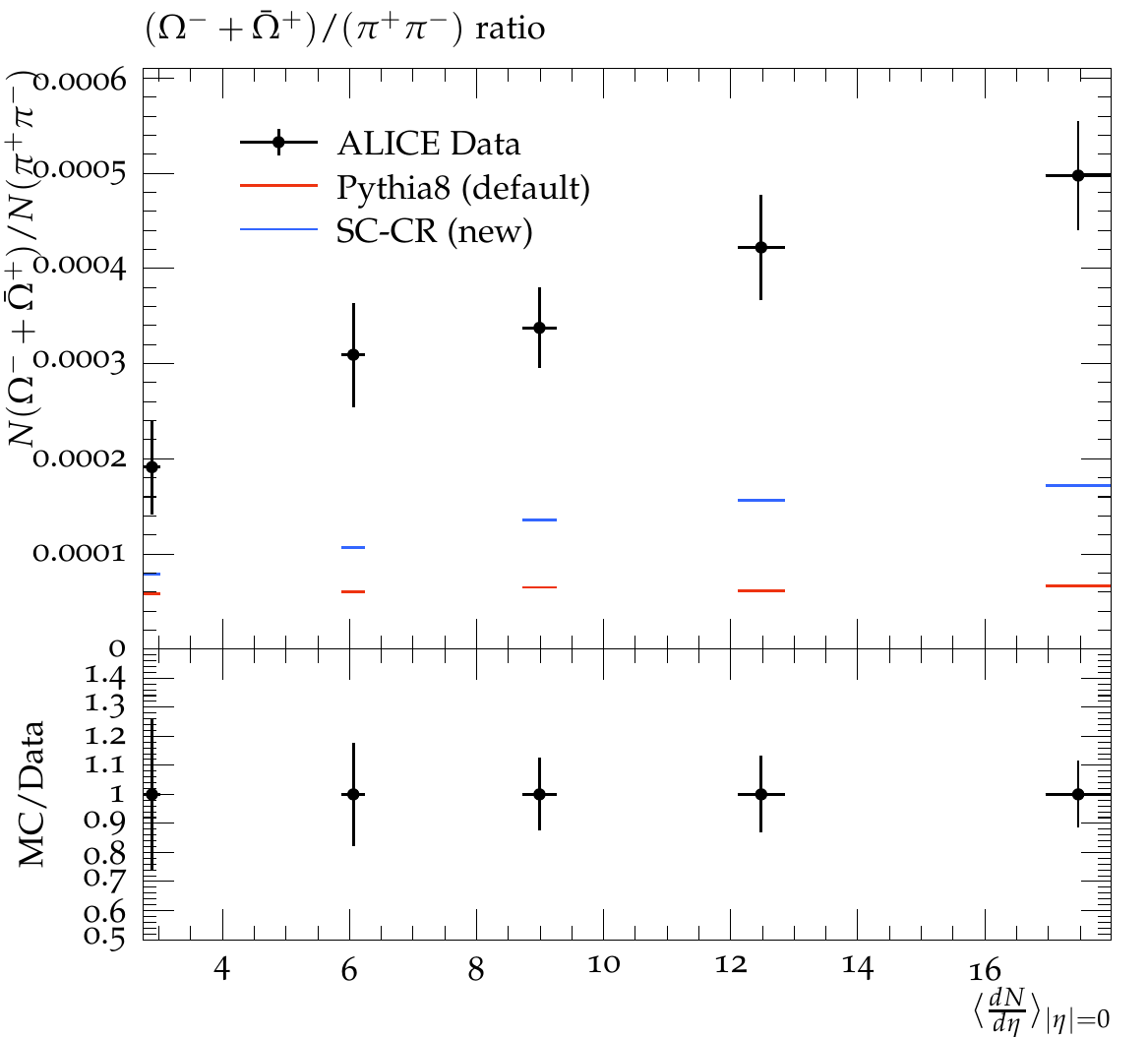}
\end{subfigure}%
\caption{From left to right the ratios of $\Xi/\pi$ and $\Omega/\pi$ is plotted against the average charged-particle multiplicity in the  
  central pseudo-rapidity. \pp collisions minimum bias events are generated 
  at $\sqrt{s}=$7~TeV and compared with ALICE \cite{ALICE:2016fzo}
  results. The lines are the same as in Figure \ref{fig:pp_mult}.}
\label{fig:pp_ratio_hyp}
\end{figure*}

Although we in this paper are mainly concerned with general
multiplicities and transverse momentum distributions, it is
interesting to also look at other effects of the QCD-CR model. In
particular the junction reconnections are interesting, as they are
known to substantially affect the baryon-to-meson ratios.

Figures \ref{fig:pp_ratio} and \ref{fig:pp_ratio_hyp} show
baryon-to-meson ratios for non-strange and strange baryons for the \pp
collision at $\sqrt{s}=$7~TeV and compared with ALICE
\cite{ALICE:2016fzo} results. The events are generated with default
\pythia set-up and for the spatially constrained QCD CR model. The
ratios of $p/\pi$, $\Lambda/\pi$, $\Lambda/K^0_s$, $\Xi/\pi$, and
$\Omega/\pi$ are produced for the combined yield of particles and
respective anti-particles. We observe that the baryons production is
enhanced, which is an expected outcome of using the QCD CR model
\cite{QCDCR}.  From the $p/\pi$ ratio plots in figure
\ref{fig:pp_ratio}, we notice that the model produces too many
protons, while the model improves the distribution of $\Lambda$,
$\Xi$, and $\Omega$ baryons. We note that in \pp the spatial
constraint is not very important for these ratios, and similar results
are expected for the standard QCD-CR model without our modifications.

The results of the $\Lambda/K^0_s$ ratio are showing agreement with
the ALICE data similar to the results obtained using the
rope-hadronization \cite{Bierlich:2022ned} model in \pythia.  It is
important to note that the CR model acts on the colour dipoles, while
the rope-hadronization acts on the Lund strings during the string
fragmentation.  Hence it will be interesting to investigate if both
models together can improve \pyt results and to what extent each of
the models will influence the hadrons yield in all three collision
systems.  At the moment from the figures \ref{fig:pp_pt} and
\ref{fig:pp_ratio} the $p_t$ distribution of the charged particles and
the proton yield both are shortcomings of the spatially constrained
QCD CR model.

\subsection{\pPb results}
\label{S:pPb}

\begin{figure}[t]
\centering
  \includegraphics[width=1\linewidth]{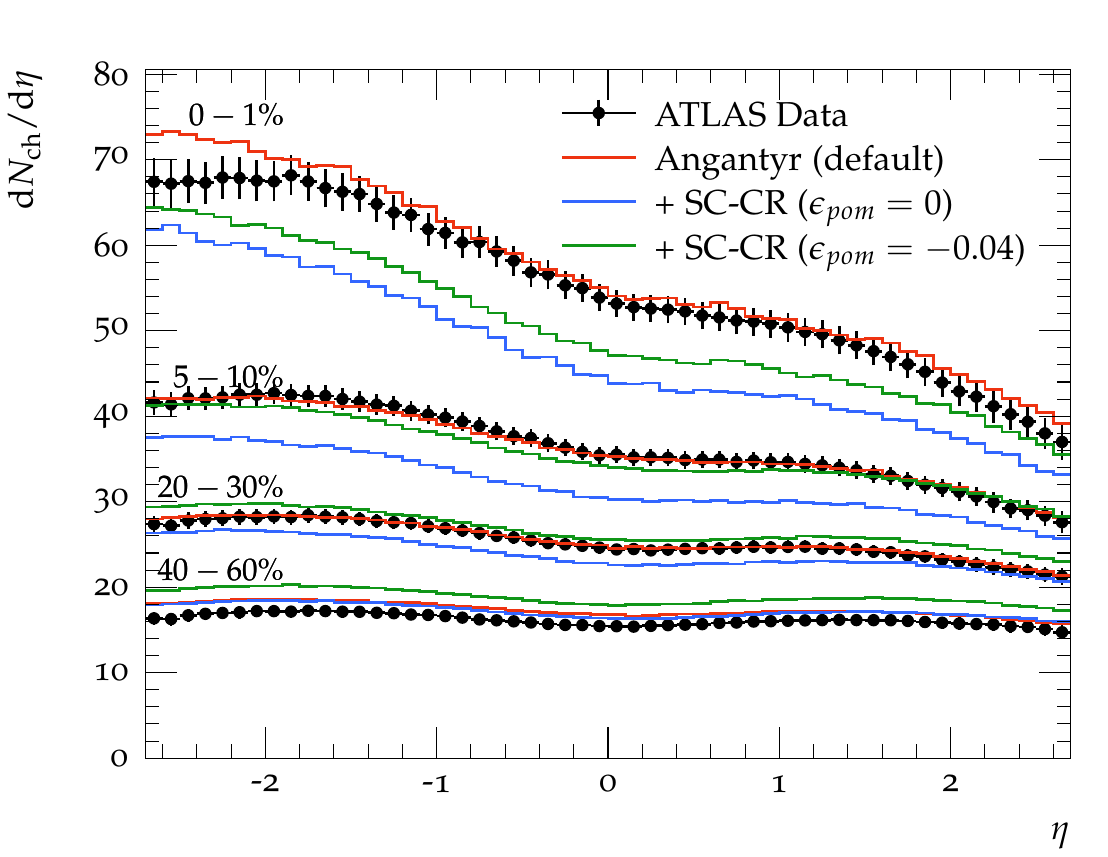}
  \caption{Average charged hadron multiplicity as a function of $\eta$ for different
    centrality bins for \pPb events are generated at
    $\sqrt{s_{NN}}=5.02$~TeV and compared with ATLAS
    \cite{pPbATLAS:2015hkr} results. The red lines are Angantyr
    default, the blue and green ones are also from Angantyr but the
    blue with the spatially constrained QCD-CR (with the new tune obtained
    for \pp collisions), and green are the same but with the tuned
    value of $\epsilon_{pom}=-0.04$. The top group of lines show the
    results from the centrality interval $0-1$\%, followed by the
    centrality intervals $5-10$\%, $20-30$\% and, at the bottom,
    $40-60$\%.}
\label{fig:pPb_multi}
\end{figure}

We show here comparisons between the Angantyr-generated average
charged particles multiplicity as a function of pseudo-rapidity in
different centrality bins for \pPb collision events at
$\sqrt{s_{NN}}=5.02$~TeV, and experimental results from ATLAS
\cite{pPbATLAS:2015hkr}. For the SC-CR we use the parameter values we
tuned to \pp\ with the heavy-ion specific parameters listed in table
\ref{tab:AA}, while the other parameters retain their values for
QCD-CR (mode-0) in table \ref{tab:table1}.  The centrality
determination is made separately for each generated dataset using the
standard Rivet \cite{Rivet} routine.\footnote{ The centrality routine
  is called \setting{ATLAS\_pPb\_Calib}, and the multiplicity analysis
  is then made with the option \setting{cent=GEN} in Rivet.}

The results are shown in Figure \ref{fig:pPb_multi}, and it is clear
that just allowing for CR between individual sub-collisions (blue
lines labelled ``+ SC-CR ($\epsilon_{pom}=0$)'') seriously degrades
the good reproduction of data for the default Angantyr setup. This is
expected as the CR necessarily will reduce the multiplicity in events
with many participating nucleons. Trying to counteract this by
decreasing $\epsilon_{pom}$ and thus increasing the multiplicity of
SND sub-events, improves the reproduction of data, but as seen in
Figure \ref{fig:pPb_multi} (green lines), for the most central events
it is not quite enough. We also see that for more peripheral events
there is a tendency to overestimate the multiplicity, and decreasing
$\epsilon_{pom}$ further to increase multiplicity in central events
would also worsen the description of peripheral events.

It should be noted here that the concept of centrality in \pPb, is a
bit complicated, and we showed in \cite{Angantyr} that any centrality
measure will not only be sensitive to the number of participating
nucleons, but will also be sensitive to multiplicity fluctuations in
individual sub-collisions, especially the most central bins. Looking
back on figure \ref{fig:pp_pt}, we see that our SC-CR tune have much
fewer fluctuations to very large multiplicities than default \pyt in
\pp\ events, and it is possible that improving this could also help
the description of very central \pPb\ events.

\begin{figure*}
\centering
\begin{subfigure}{.5\textwidth}
  \includegraphics[width=1\linewidth]{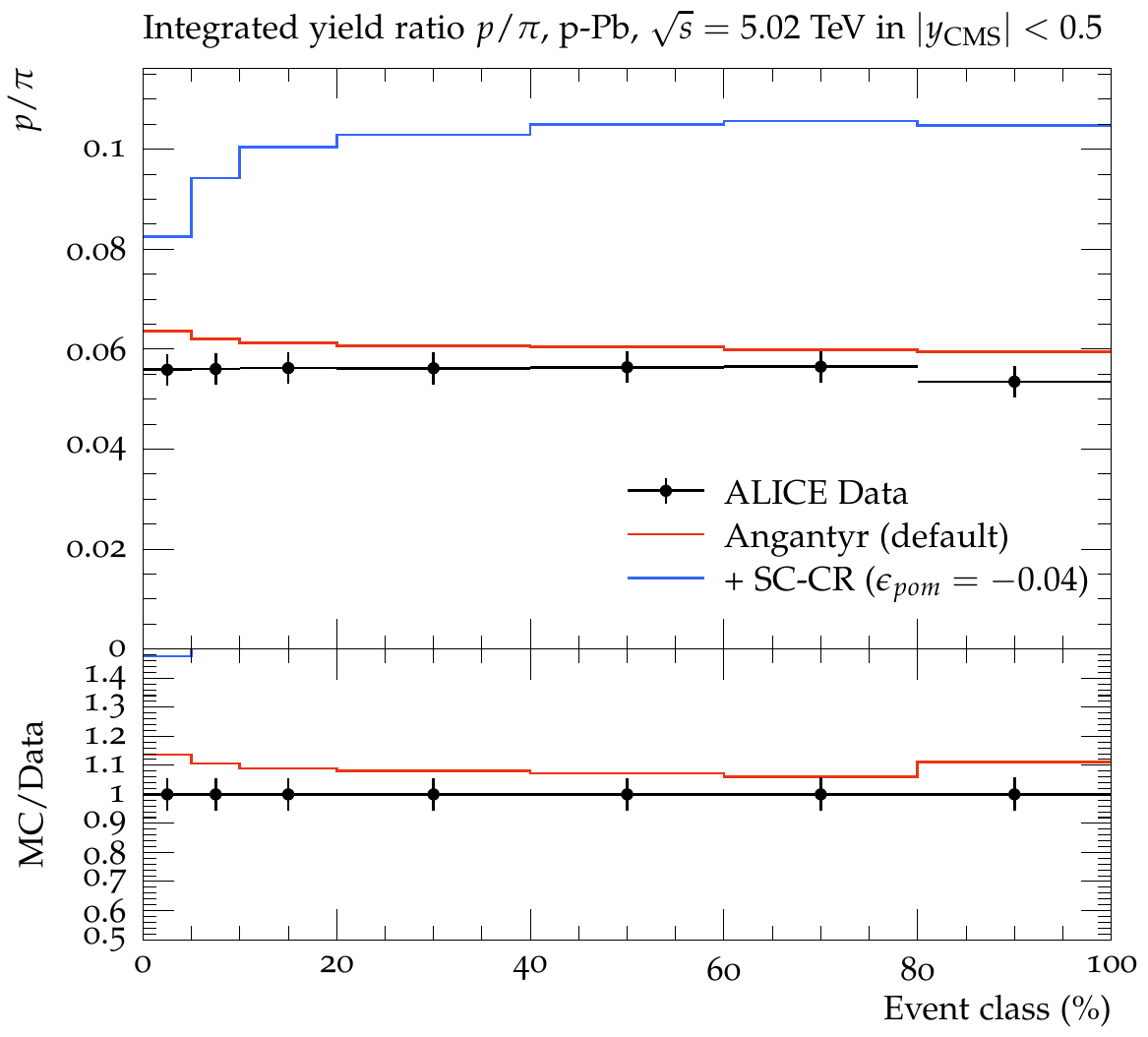}
\end{subfigure}%
\begin{subfigure}{.5\textwidth}
  \includegraphics[width=1\linewidth]{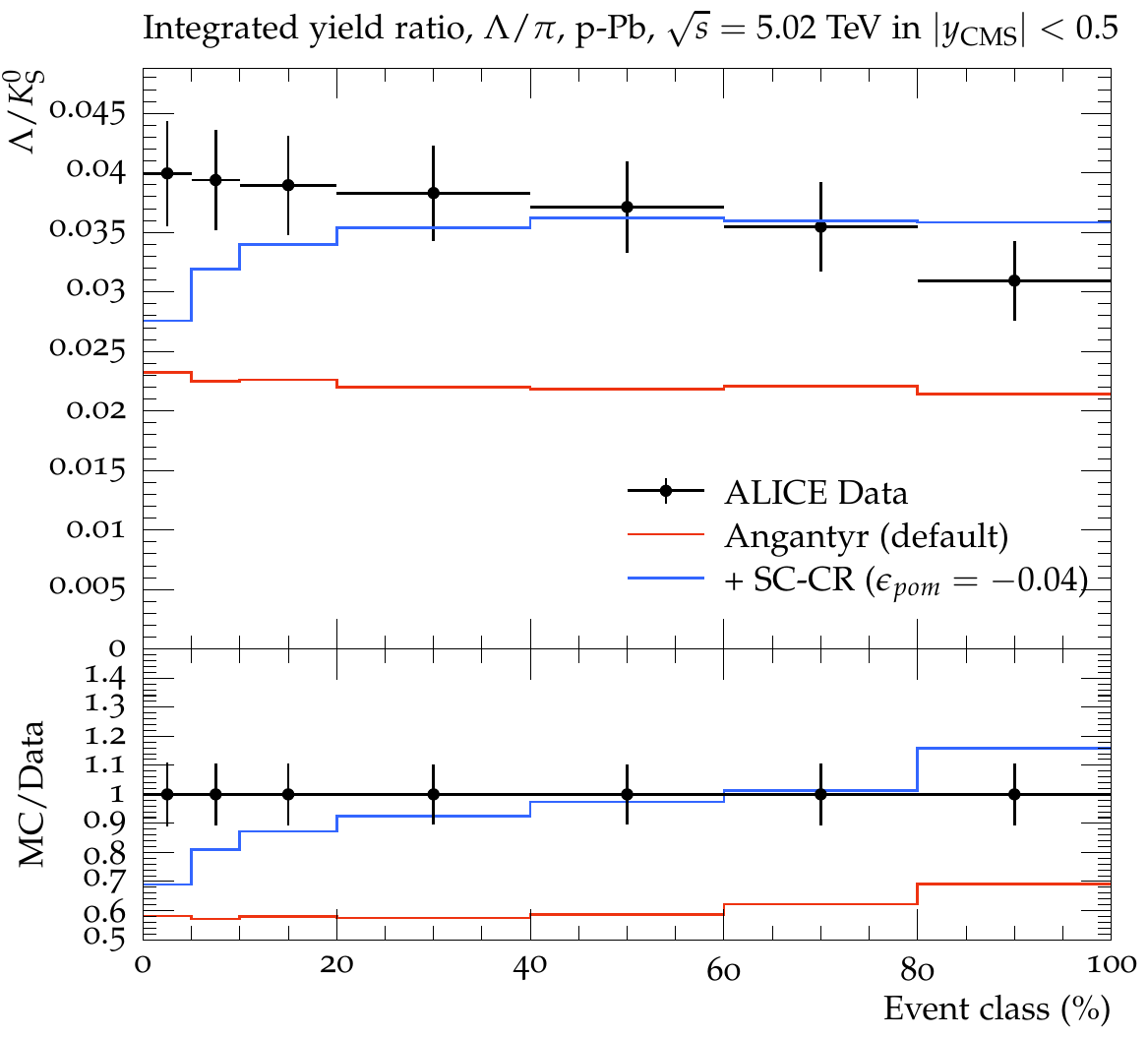}
\end{subfigure}%
\caption{The ratios of $p/\pi$ (left) and $\Lambda/\pi$ (right) are plotted 
for different centrality for \pPb\ collision events at $\sqrt{s}=$5.02~TeV and 
compared with ALICE \cite{ALICE:2013ratio} results. The red line is generated 
with the default Angantyr settings, while the blue line is SC-CR with the tune
obtained for \pA\ collisions.}
\label{fig:ppb_ratio}
\end{figure*}

Figure \ref{fig:ppb_ratio} shows the baryon-to-meson ratio for protons and $\Lambda$ baryons in \pPb collision events $\sqrt{s_{NN}}=5.02$~TeV, and the results are compared with ALICE data \cite{ALICE:2013ratio}.
We can notice that the results follow the trend of baryon-to-meson ratio in Figure \ref{fig:pp_ratio}, the SC-CR model produces too many protons irrespective of the event centrality, while the $\Lambda/\pi$ distribution is improved compared to Angantyr (default).

\subsection{\PbPb results}
\label{S:AA}

\begin{figure*}[t]
\begin{subfigure}{.5\textwidth}
  \includegraphics[width=1\linewidth]{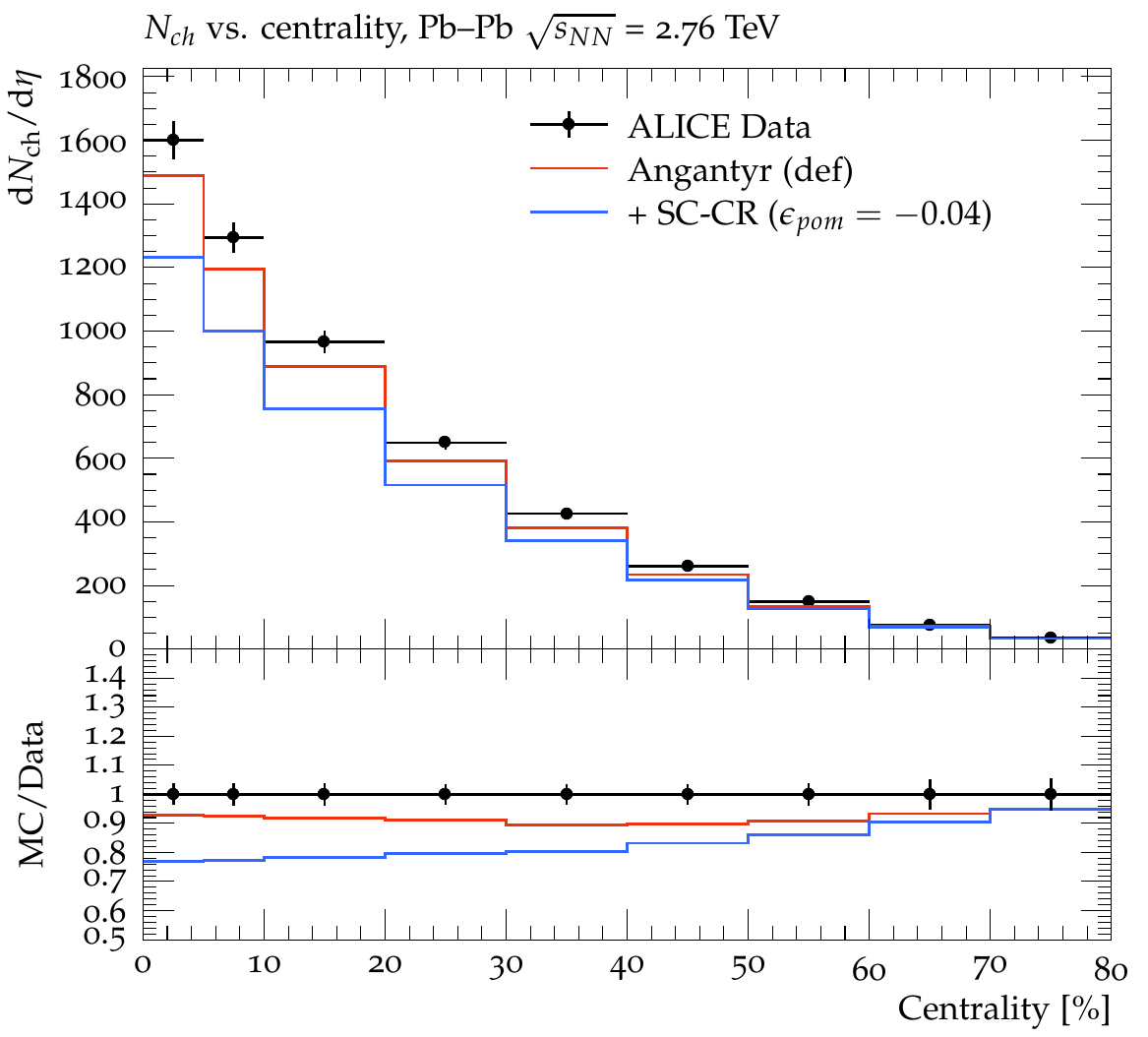}
\end{subfigure}%
\begin{subfigure}{.5\textwidth}
  \includegraphics[width=1\linewidth]{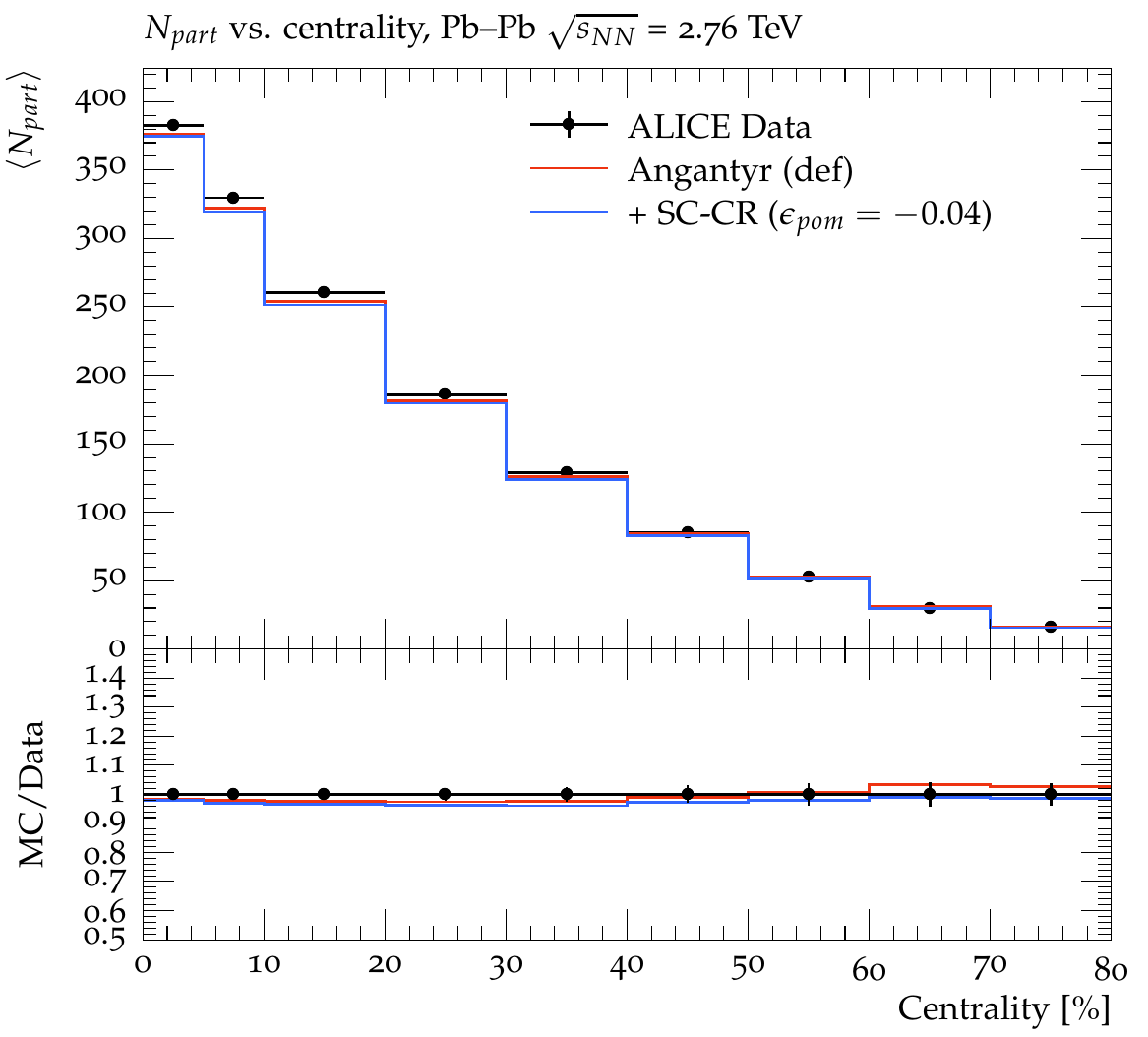}
\end{subfigure}
\caption{$\textit{Left}$: Central ($\mid \eta\mid <0.5$) charged multiplicity as a
  function of centrality for \PbPb collisions at
  $\sqrt{s_{NN}}=2.76$~TeV.  The red line is generated with the
  default Angantyr settings, while the blue line is SC-CR with the
  tune obtained for \pA\ collisions. The data is from ALICE
  \cite{ALICE:2010mlf}. $\textit{Right}$: number of participating nucleons,
  $N_{part}$, as a function of centrality for the same event samples,
  compared with Glauber-model calculations from ALICE
  \cite{ALICE:2010mlf}.}
\label{fig:PbPb_multi_alice}
\end{figure*}

\PbPb events at $\sqrt{s_{NN}}=2.76$~TeV were generated using the
Angantyr model with its default setup and the SC-CR setup (table
\ref{tab:AA}).  The simulation results are compared with ALICE
\cite{ALICE:2010mlf} in figure \ref{fig:PbPb_multi_alice} (left) for
the central ($\mid \eta \mid <0.5$) charged multiplicity in different
centrality bins\footnote{Again the centrality bins are calculated on
  the generated data, this time using the
  \setting{ALICE\_2015\_PBPBCentrality} analysis in Rivet}. We see
here the same tendency as in \pPb, that the multiplicity is reduced
for central events when including CR between sub-collisions, while for
more peripheral events it is less affected. The effect is roughly the
same, $\sim20$\%, for the most central events. Even though the
multiplicity in \PbPb\ is much higher, the sub-collisions are more
spread out in impact parameter, and the spatial constraint, \ADS,
therefore severely restricts the effect of CR.

It should be noted that the effects on the centrality binning of
fluctuations in sub-collision multiplicity are not so important in
\PbPb, compared to \pPb. This is reflected in the average number of
participating nucleons as a function of centrality obtained from a
Glauber model, shown in figure \ref{fig:PbPb_multi_alice}
(right). Here we see that the agreement between Angantyr, with and
without SC-CR agrees very well with each other and with the number
obtained from the data.

\begin{figure*}[t]
\centering
\begin{subfigure}{.5\textwidth}
  \includegraphics[width=1\linewidth]{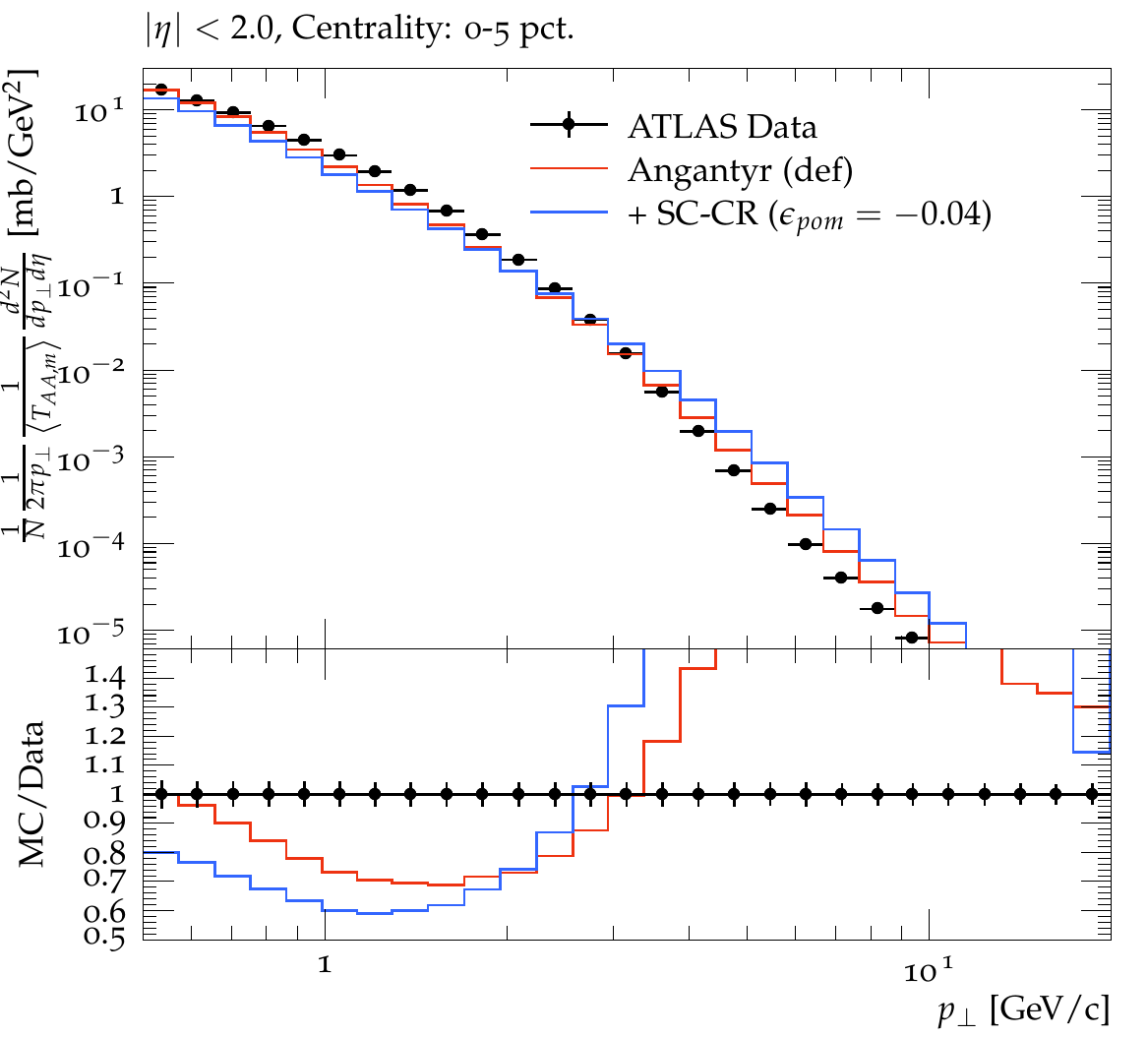}
\end{subfigure}%
\begin{subfigure}{.5\textwidth}
  \includegraphics[width=1\linewidth]{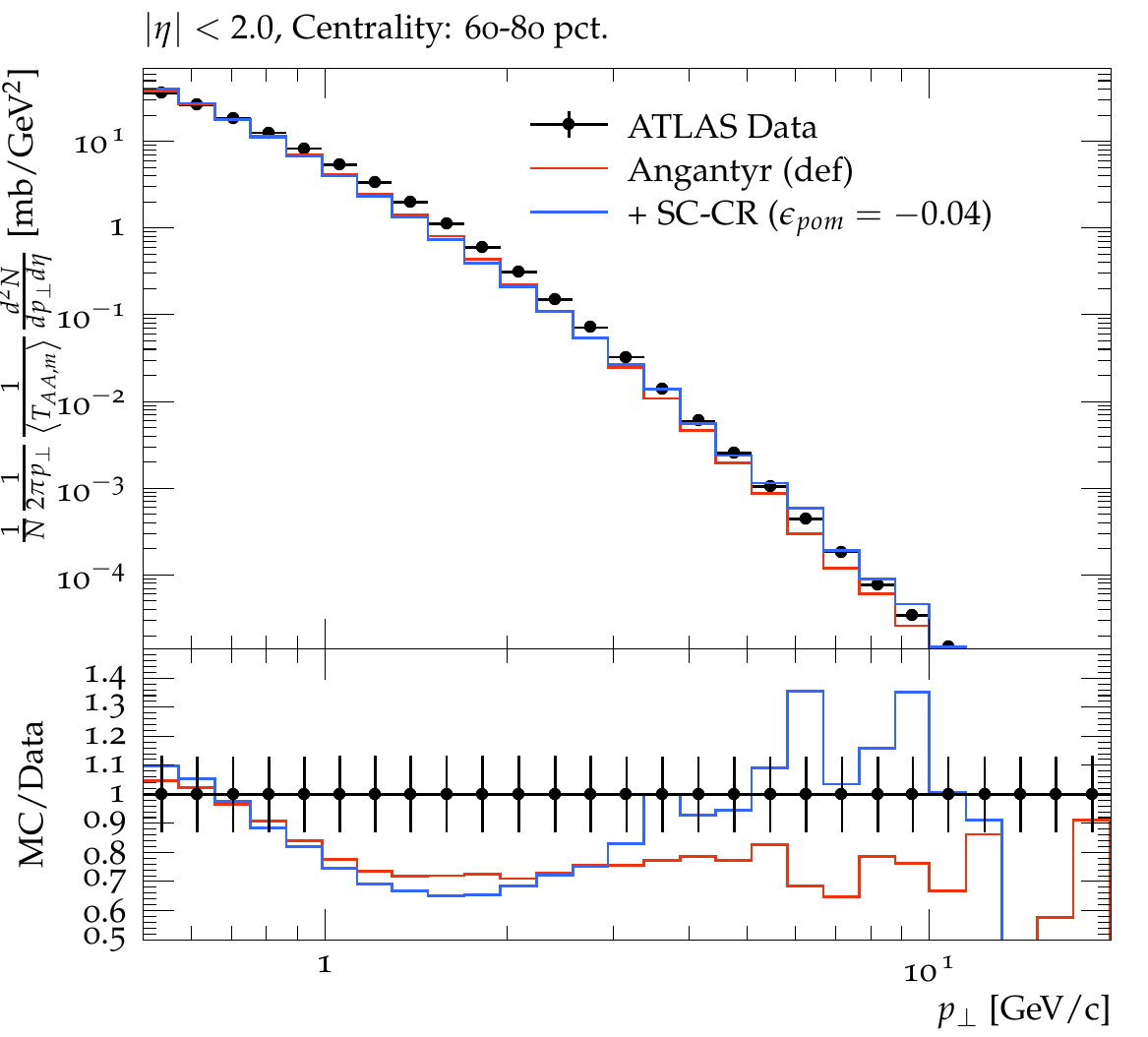}
\end{subfigure}
\caption{Transverse momentum distribution of charged particles in
  different centrality bins for \PbPb collisions, generated at
  $\sqrt{s_{NN}}=2.76$~TeV and compared with ATLAS
  \cite{PbPbATLAS:2015qmb} results. The lines are the same as in
  figure \ref{fig:PbPb_multi_alice}.}
\label{fig:PbPb_pt_atlas}
\end{figure*}

We noted in \cite{Angantyr} that, while the default Angantyr gives a
reasonable description of the multiplicity in \PbPb\ data, other
observables were not as well reproduced. In particular this applies to
$p_\perp$ spectra, which we show here in figure
\ref{fig:PbPb_pt_atlas} compared with ATLAS \cite{PbPbATLAS:2015qmb}
for central and peripheral events. Considering that we already
in figure \ref{fig:pp_pt} showed that SC-CR degrades the description of the
$p_\perp$ distribution in \pp,\ it would be unlikely that it would be
better in \PbPb. Indeed we see in the figure that adding global colour
reconnection in \PbPb, rather makes the description of data worse
in central collisions. For peripheral collisions, one could argue that
there is an improvement in the high-$p_\perp$ tail, but the overall
performance is still not very good.

So far for \pp and \pPb, we have shown proton-to-pion and some of the hyperons-to-pion ratios.
For \PbPb we want to refrain from showing any baryon-to-meson results.
The SC-CR model fails to reproduce the charged multiplicity for many centrality bins in \PbPb collision events.
Hence the model's agreement or disagreement with the experimental data of baryon-to-meson ratios will be irrelevant at this stage.
We want to improve the model description for the overall charged multiplicity and the $p_t$ distribution of the produced particles before we test the model efficiency against the different identified particle yields in \AA collisions.

The SC-CR model increases the baryon production in \pp and \pPb collision systems.
The result of strangeness enhancement in the strange baryons sector without any assumption of thermalised medium opens the possibility for an alternative explanation for the observed strangeness enhancement in heavy-ion collisions. 
The results from heavy hyperons are much below the experimental observations (figure \ref{fig:pp_ratio_hyp}, but we should note that these results are producing a similar trend of strangeness enhancement to that of rope-hadronization \cite{Bierlich:2022ned}.
We hope that in future combining these two models will improve the \pythia/Angantyr model efficiency in reproducing the strange hadrons yield.

\section{Discussion and outlook}
\label{S:discussion}

We have here presented a first attempt to extend the concept of colour
reconnections to apply also between sub-collisions in heavy ion
events. This was done by modifying the QCD colour reconnection model
in \pythia to limit the spatial distance between dipoles that are
allowed to reconnect. This was done with a simple cutoff in impact
parameter distance, \ADS, for which we found a value $0.5$~fm, to give
reasonable results.

This is clearly a quite naive approach, and the aim is mainly to
understand the phenomenology of allowing inter-sub-collision CR in
heavy ion collisions. There are many paths to improve our simple
model. It is not unnatural, \eg, to let the value of \ADS depend on
the local density of dipoles, and/or the transverse momentum of the
partons involved in the reconnections. The role of $m_0$ also needs
to be studied more, and this parameter could also be allowed to depend
on such local features of the event. There are also other reconnection
models that can be considered for use in heavy ion collisions, such as
the perturbative swing model in Ref.\ \cite{rope1, Lonnblad:1995yk}.

Our results are not perfect, but we still feel that they are
encouraging. Introducing the cutoff makes the reproduction of \pp\
minimum bias data worse than in default \pythia, but not very much
worse than what is obtained with the QCD-CR model. For \pA\ collisions
we get a reduced multiplicity, as expected, but we find that it can
be compensated by modifying the treatment of secondary non-diffractive
sub-collisions in the Angantyr model. Finally, we find that when
extrapolating to full \AA\ events we again obtain a too low
multiplicity for central events, but the reduction is typically below
20\%.

It should be remembered here that there are many other effects to
consider, both in high-multiplicity \pp\ collisions and in heavy ion
collisions. In our group, we have considered so-called string shoving
\cite{shoving,shoving2} and rope hadronization \cite{rope1}, as well
as hadronic rescattering \cite{hadrescat}. Especially the latter would
be interesting study together with colour reconnection in heavy ion
collisions, since it has been found that rescattering typically
increases the multiplicity for central collisions \cite{hardrescat2}.

The QCD-CR model has shown enhanced production of the baryons compared
to the default MPI-based CR model, which is purely a byproduct of
junction systems.  Hence, there may be some effects similar to
strangeness enhancement due to the QCD-CR model in the baryon sector
\cite{strangeCR}.  The QCD-CR model has shown effects similar to
collectivity \cite{collectivity}.  This work will also enable us to
study and investigate the contribution of different event simulation
stages like CR, string interactions (shoving and rope), and hadronic
rescattering on the various final state observables in \pp, \pA, and
\AA collisions.  In future work, we will show the results of the new
implementation for the flavour production and collectivity-like
behaviour in all three collision systems, namely \pp, \pA, and \AA.

A top priority for future work is therefore to develop our
implementation to allow the study of the simultaneous effects of all
these models. Only then one would be able to do a proper tuning of the
involved parameters. In doing so we would again follow the procedure
used here, \ie, first tune to minimum bias \pp\ observables of
multiplicities and transverse momentum, and then adjust parameters
that also depends on inter-sub-event effect by tuning to similar
observables in \pA\ in order to be able to have a parameter-free
extrapolation to \AA\ events.

\bmhead{Acknowledgments}

We thank G{\"o}sta Gustafson, Torbj{\"o}rn Sj{\"o}strand and Christian Bierlich
for interesting discussions and important input to this work.

This work was funded in part by the Knut and Alice Wallenberg
foundation, contract number 2017.0036, Swedish Research Council,
contracts numbers 2016-03291 and 2020-04869, in part by the European
Research Council (ERC) under the European Union’s Horizon 2020
research and innovation programme, grant agreement No 668679, and in
part by the MCnetITN3 H2020 Marie Curie Initial Training Network,
contract 722104.

\begin{appendices}
\section{Junction Fragmentation}
\label{S:jun_frag}

The hadronization in \pyt is done through Lund string fragmentation.
\pyt has two methods for hadron production; string fragmentation and
mini string fragmentation.  Both are fundamentally based on the same
mechanism, but the latter is an approximation for short strings
(mini-strings), which becomes more important with many reconnections
in HI events.  Depending on the invariant mass of the colour singlet
string/junction system, \pyt's algorithm selects one of the two
methods to produce primary hadrons.

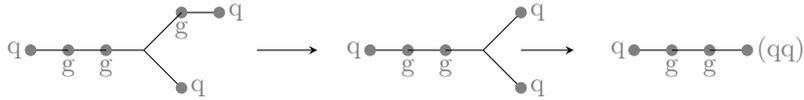
\begin{figure*}[h!]
\centering
\begin{tikzpicture}
\draw (-2.5,-1) -- (-3,-1);
\filldraw [gray] (-2.5,-1) circle (2pt) node[right]{q};
\filldraw [gray] (-3,-1) circle (2pt) node[below]{g};
\draw (-3,-1) -- (-3.5,-1.5);
\draw (-3.5,-1.5) -- (-3,-2);
\filldraw [gray] (-3,-2) circle (2pt) node[right]{q};
\draw (-3.5,-1.5) -- (-4,-1.5);
\filldraw [gray] (-4,-1.5) circle (2pt) node[below]{g};
\draw (-4,-1.5) -- (-4.5,-1.5);
\filldraw [gray] (-4.5,-1.5) circle (2pt) node[below]{g};
\draw (-4.5,-1.5) -- (-5,-1.5);
\filldraw [gray] (-5,-1.5) circle (2pt) node[left]{q};

\draw [-stealth](-2,-1.5) -- (-1.2,-1.5);
\draw (-0.5,-1.5) -- (0,-1.5);
\filldraw [gray] (-0.5,-1.5) circle (2pt) node[left]{q};
\filldraw [gray] (0,-1.5) circle (2pt) node[below]{g};
\draw (0,-1.5) -- (0.5,-1.5);
\filldraw [gray] (0.5,-1.5) circle (2pt) node[below]{g};
\draw (0.5,-1.5) -- (1,-1.5);

\draw (1,-1.5) -- (1.5,-2);
\filldraw [gray] (1.5,-2) circle (2pt) node[right]{q};
\draw (1,-1.5) -- (1.5,-1);
\filldraw [gray] (1.5,-1) circle (2pt) node[right]{q};

\draw [-stealth](1.5,-1.5) -- (2.2,-1.5);

\draw (3,-1.5) -- (3.5,-1.5);
\filldraw [gray] (3,-1.5) circle (2pt) node[left]{q};
\filldraw [gray] (3.5,-1.5) circle (2pt) node[below]{g};
\draw (3.5,-1.5) -- (4,-1.5);
\filldraw [gray] (4,-1.5) circle (2pt) node[below]{g};
\draw (4,-1.5) -- (4.5,-1.5);
\filldraw [gray] (4.5,-1.5) circle (2pt) node[right]{(qq)};
\end{tikzpicture}
\caption{Junction fragmentation steps. $\textit{Left}$: A colour singlet junction
  system. In this example, the lowest energy leg is the one with a
  quark directly attached to the junction. The highest energy leg is
  the one with the longest chain of dipoles. $\textit{Centre}$: The junction
  system after both the low-energy legs are fragmented and left with a
  quark directly connected to the junction. The highest energy leg is
  intact. $\textit{Right}$: The final state of the junction system is
  shown. Here, the quarks of the two low-energy junction legs are
  combined to form a diquark (represented as $(qq)$), and it is
  attached at the end of the highest-energy leg. The remaining
  junction system is now replaced with a single string, and it will
  fragment according to the normal string fragmentation.}
\label{fig:junfrag}
\end{figure*}

String fragmentation fragments long strings.  For the junction
systems, the algorithm fragments two lower energy junction legs first, and
it starts from the lowest energy leg.  The two low-energy junction
legs are fragmented until every leg is left with a parton directly
connected to the junction.  Later, the two partons from these low-energy legs are
combined to form a diquark (or anti-diquark).  The diquark is then
treated as one end of the remaining highest energy junction leg.  At
this stage, the junction system no longer exists, and the last
junction leg with a diquark at one end undergoes further string
fragmentation as a normal string.  Figure \ref{fig:junfrag} shows the
progress of the junction break-ups, formation of a diquark, and
transformation of the junction system into a string system.

Mini-string fragmentation is used to hadronize short strings.  It
produces one or two hadrons depending on the energy in the string, and
low-energy junction systems are not treated in the mini-string
fragmentation.  Prior to this work, if low-energy junction systems are
found then such an event is aborted and a new event is generated.
When we merged multiple \pp-like events at the parton level in a HI
event and tried to perform CR in the entire event, we observed
enhancement in the number of junction systems, and many of them are
below the invariant mass cut-off for string fragmentation.  All those
junction systems have to be fragmented within a mini-string
fragmentation module, otherwise, the majority of HI events are aborted
due to one or more untreated colour singlet systems in the
hadronization stage.  Hence, we developed two new fragmentation
functions: "MiniJunct2Hadrons" and "MiniJunct2Baryons", in the
mini-string fragmentation.

Before we provide an overview of the technical details of the two new
functions, we like to discuss some additional constraints we
introduced inside the QCD-CR model.  We observe that most of these
junction systems in \pA\ and \AA systems, have diquarks in the junction
legs.  Now, as we mentioned earlier, in the string fragmentation the
two lower energy legs of a junction system are fragmented and the
remainder is merged to form a string connecting the diquark with the
highest energy junction leg.  But in the case of the junction being
connected to a diquark, it often occurs that either both or one of the
low-energy legs is left with a diquark.  \pyt can not merge such
junction legs, where one or both legs have a diquark.

Sometimes, the highest energy leg also has a diquark at the end.  Now,
after merging two low-energy legs into a diquark, the string has a
diquark at both ends.  Although string fragmentation is a
probabilistic mechanism and the strings with two diquarks may
fragment successfully, sometimes they fail due to the last string
piece being left with a diquark at both ends.  The last piece can not be hadronized as a tetra-quark system at this moment,
although that could in principle be considered.  We introduce an
additional attempt by forwarding such a failed string fragmentation to
mini-string fragmentation, where the string/junction system is forced
to produce one or two hadrons.

\begin{figure}
\centering
  \includegraphics[width=1\linewidth]{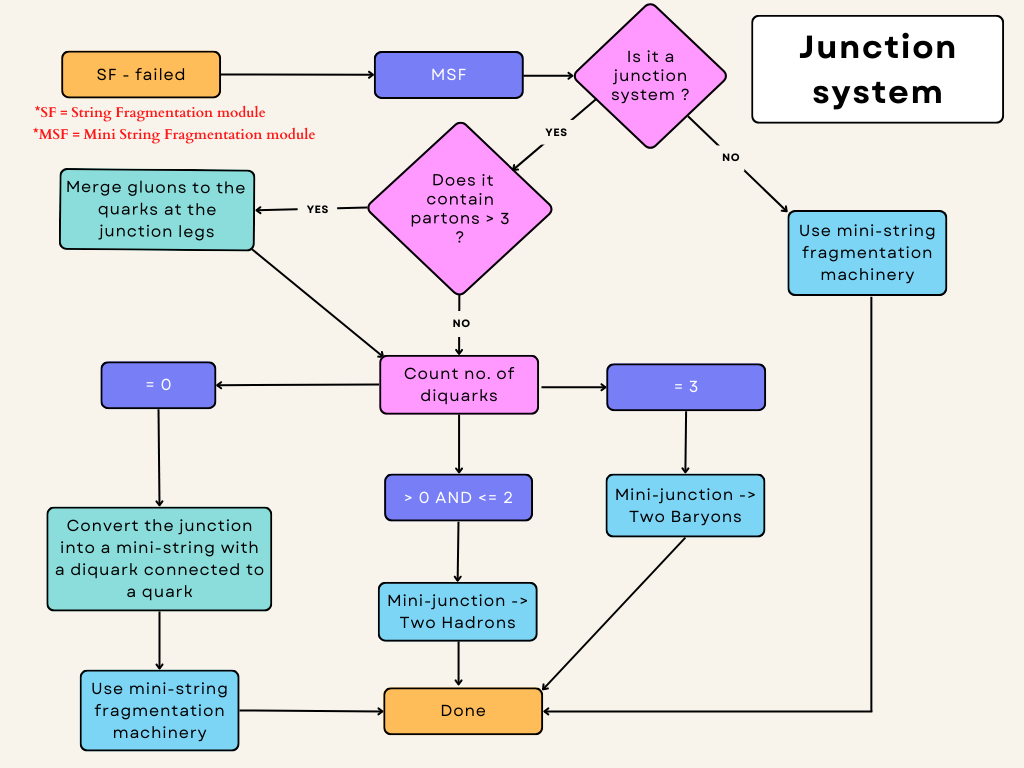}
\caption{Overview of the steps taken if the string fragmentation module fails to produce hadrons, and the handling of junction systems in the mini-string fragmentation to produce primary hadrons. The flow chart shows examples of cases with quarks and diquarks, but the junctions can also have anti-quarks and anti-diquarks.}
\label{fig:MJO}
\end{figure}

To avoid recurring failures in the string fragmentation due to
junction systems containing diquarks in \pA\ and \AA collisions, we do
not allow dipoles with a diquark to undergo CR.  This constraint is
provided with a Boolean flag, so future users can test the options
with or without allowing CR for diquarks containing dipoles.  In \pp
collisions, another source of diquarks prior to string fragmentation
is beam remnants.  \pyt allows users to choose if the beam remnants
may form a junction or a diquark with the remaining valance quarks of
the beam remnant.  Here, we choose junction formation to reduce the
preexisting number of diquarks in the event during the CR.  We
continue to use this setup not only for \pp collisions but also for
\pA\ and \AA collisions.

Now, let's go back to the junctions in mini-string fragmentation.  The
first step is to check if the junction contains any gluons or not.  If
any of the junction legs have gluons, we reduce the junction system
size by absorbing the gluons into the diquark or quark of the
respective junction legs.  The idea here is to simplify the next steps
of producing hadrons via "junction collapse".  We reduce the complex
junction system to a simple junction system, where every leg contains
only a quark or a diquark or respective anti-particle at its end.
Now, we calculate the number of diquarks in the junction system.  As
shown in Figure \ref{fig:MJO}, the junction can have up to three
diquarks, when all the junction legs have a diquark.  Depending on the
number of diquarks, we have three possible outcomes for the junction
to collapse and produce primary hadrons.

Case A (All the legs have quarks):

In this case, we follow the steps similar to the last stage of junction fragmentation in the string fragmentation mechanism.
We merge the two lowest energy legs to form a diquark.
Then we reduce the junction into a mini-string containing a diquark and a quark.
After that, we use the existing mini string fragmentation functions to produce primary hadrons.

Case B (The junction legs have a maximum of two diquarks):

Here, depending on whether the junction has one or two diquarks we
collapse it to produce two mesons, or one baryon and one meson.  We
observed that the junctions containing one diquark have anti-quarks on
the remaining legs and if it's an anti-diquark then quarks. Similarly, in the case of two diquarks, then the remaining leg has an anti-quark, and if two anti-diquarks then a quark. Hence,
we treat the junction as a single entity containing a given number of
diquarks, anti-diquarks, quarks and anti-quarks, and we produce
hadrons accordingly.

Case B.1 (One diquark, and two anti-quarks):

Here, we break the diquark into two quarks.
Then we pair each quark with a randomly selected anti-quark from the two anti-quarks.
We check if the produced mesons have the correct quark flavours or not.
Then we check if the sum of their masses is less than the invariant mass of the junction system or not.
If the sum of the masses is below the junction system's invariant mass, then the mesons are assigned momenta as if a mother particle decays into the two daughter particles. 
If it fails  then we repeat the process several times.
If all attempts fail, the function returns a message about failed attempts, and the event is aborted.

Case B.2 (Two diquarks and one anti-quark):

Here, a diquark with the lowest invariant mass is broken into two quarks.
One of the quarks is randomly selected and assigned to the other diquark, and the other quark is paired with the anti-quark.
Again we check the flavours and compare the sum of the masses of the baryon and the meson with the junction system mass.
If the sum of the masses is below the junction system's invariant mass then the hadrons are assigned momenta as if a mother particle decays into the two daughter particles. 
If it fails then we repeat the process several times.
If all attempts fail, the function returns a message about failed attempts, and the event is aborted.

Case C (All legs have diquarks):

In this case, we have in total six quarks, and we collapse the junction to produce two baryons.
We follow the procedure similar to case B.2, the diquark with the lowest invariant mass is split into two quarks.
One of the quarks is randomly assigned to one of the remaining two diquarks.
The rest of the steps are the same.

\end{appendices}

\bibliography{articlemain}


\end{document}